\PassOptionsToPackage{numbers,sort&compress}{natbib}
\documentclass{article}

\usepackage[preprint]{neurips_2026}

\usepackage[utf8]{inputenc} 
\usepackage[T1]{fontenc}    
\usepackage{hyperref}       
\usepackage{url}            
\usepackage{booktabs}       
\usepackage{amsfonts}       
\usepackage{nicefrac}       
\usepackage{microtype}      
\usepackage{xcolor}         
\usepackage{graphicx}
\usepackage{amsmath}
\usepackage{multirow}
\usepackage{subcaption}
\usepackage{tabularx}
\usepackage{booktabs}
\usepackage{longtable}
\usepackage{placeins}

\title{MatFormBench: A Benchmarking Evaluation Framework for Target-Driven Materials Formulation}

\author{%
  Linhan Wu \\
  DeepVerse\\
  \texttt{linhan@deepverse.tech} \\
  \And
  Chenxi Wang \\
  DeepVerse\\
  \texttt{tracy@deepverse.tech} \\
  \And
  Chuhan Yang \\
  DeepVerse\\
  \texttt{chuhan@deepverse.tech} \\
  \And
  Zhengwei Yang \\
  DeepVerse\\
  \texttt{zhw\_yann@shu.edu.cn} \\
  \And
  Yuyang Liu\thanks{Corresponding author.} \\
  DeepVerse\\
  \texttt{yuyang@deepverse.tech} \\
}

\begin{document}

\maketitle

\begin{abstract}
Inverse design of materials has significantly advanced target-driven formulation optimization, yet existing materials machine learning benchmarks remain limited to forward property prediction, failing to systematically evaluate inverse optimization and generation algorithms—a critical gap that hinders the progress of target-driven materials design. To address this limitation, we propose MatFormBench, a novel benchmarking ecosystem tailored to evaluate and guide generative strategies for target-driven formulation. MatFormBench integrates a physics-driven formulation generation scheme to generate synthetic samples that faithfully emulate realistic materials structure-property response relationships, complemented by five escalating difficulty levels to quantify the complexity of these relationships. To rigorously assess algorithm performance, we further propose MatFormScore, a multi-dimensional metric that comprehensively quantifies performance across five critical axes: target success, search efficiency, exploratory capacity, robustness, and stability. We validate MatFormBench by evaluating 39 diverse inverse design algorithms, covering classical surrogate-assisted black-box search, state-of-the-art deep generative models, and increasingly popular Large Language Model (LLM)-based recommendation strategies. Across 1170 standardized algorithm-task evaluations, diffusion-based models demonstrate the strongest overall performance, while Variational Autoencoder (VAE)-based and Genetic Algorithm (GA)-based methods exhibit distinct advantages in specific scenarios. By establishing a unified evaluation standard for target-driven materials formulation, MatFormBench enables reproducible benchmarking, principled algorithm comparison, and diagnostic analysis of inverse design strategies—providing a foundational tool for advancing materials inverse design. 
\end{abstract}

\section{Introduction}

In AI for Science (AI4Science, AI4S), inverse design of materials is shifting from proof-of-concept demonstrations toward practical target-driven formulation optimization \cite{zunger2018inverse,batra2021emerging,lee2023inverse,cheng2026ai}. Unlike forward property prediction or candidate screening, inverse design aims to generate feasible materials or formulations that satisfy prescribed properties under explicit constraints and limited experimental or computational budgets \cite{sanchez2018inverse,gomez2018automatic,lookman2019active,tran2024polymer}. This transition has been enabled by advances in machine-learning structure--property modeling and large-scale materials databases \cite{jain2013materials,ward2016general}, surrogate-assisted and active optimization \cite{xue2016accelerated,lookman2019active}, Bayesian optimization \cite{shahriari2016bayesian,frazier2018tutorial}, evolutionary and genetic algorithms \cite{jennings2019genetic}, deep generative models such as variational autoencoders and generative adversarial networks \cite{kingma2014auto,goodfellow2014generative}, diffusion-based generative modeling \cite{ho2020denoising,song2021score,karras2022edm}, and large language model-based scientific recommendation or autonomous discovery systems \cite{bran2024chemcrow,boiko2023coscientist}.

Despite rapid methodological progress, the evaluation of inverse design algorithms remains fragmented. Existing materials machine learning benchmarks mainly address forward prediction, stability screening, molecular generation, or structural generation, rather than inverse optimization and formulation generation. Matbench provides standardized tasks for supervised materials property prediction, while Matbench Discovery focuses on stability prediction and screening-oriented discovery \cite{dunn2020matbench,riebesell2023matbenchdiscovery}. For molecular benchmarks, QM9 and MoleculeNet support molecular property prediction and representation learning \cite{ramakrishnan2014qm9,wu2018moleculenet}, whereas GuacaMol and MOSES focus on goal-directed molecular generation and distribution learning \cite{brown2019guacamol,polykovskiy2020moses}. More recently, the Material Generation Benchmark (MGB) extended standardized evaluation to crystal structure prediction, de novo material generation, MOF generation, and out-of-distribution generalization \cite{yan2025mgb}. However, these benchmarks are not designed to assess whether algorithms can actively search and generate target-satisfying formulations under controlled oracle budgets. Consequently, current comparisons often depend on inconsistent datasets, design spaces, oracle assumptions, budgets, and metrics, making it difficult to identify which algorithms are suitable for which formulation regimes.

This limitation is critical because inverse design for materials formulation is structurally heterogeneous. Realistic formulation spaces can be high-dimensional, mixed-type, nonconvex, multimodal, and sparsely feasible, while oracle evaluations may be noisy, costly, or invalid for failed candidates \cite{lee2023inverse,cheng2026ai}. Thus, a useful benchmark should not only report aggregate performance, but also diagnose how algorithms behave under different difficulty regimes and why they fail.

In this work, we propose MatFormBench, a benchmarking evaluation framework for target-driven materials formulation. MatFormBench constructs physics-driven controllable oracles and organizes synthetic formulation tasks into five escalating difficulty levels, covering smooth responses, variable coupling, local discontinuities, multimodal landscapes, and globally constrained sparse feasibility. It further standardizes closed-loop oracle access and introduces MatFormScore, a multi-axis metric integrating target success, search efficiency, exploratory capacity, robustness, and stability. Through \(1170\) standardized algorithm-task evaluations over \(39\) inverse design algorithms, MatFormBench establishes a reproducible and diagnostic evaluation standard for comparing inverse design strategies and analyzing their regime-specific strengths and failure modes. Code and project page of MatFormBench are available at \url{https://github.com/DeepVerse/MatFormBench} and \url{https://matformbench.deepverse.tech}.

\section{Related Work}

\paragraph{Materials machine learning benchmarks.}
Benchmarking has become a standard practice for evaluating materials machine learning models. Large-scale materials repositories, including the Materials Project~\cite{jain2013materials}, AFLOW~\cite{curtarolo2013aflow}, OQMD~\cite{kirklin2015oqmd}, and NOMAD~\cite{draxl2019nomad}, provide the data foundation for supervised property modeling. On top of these resources, Matbench~\cite{dunn2020matbench} defines standardized regression and classification tasks for materials property prediction, while Matbench Discovery~\cite{riebesell2023matbenchdiscovery} focuses on crystal stability prediction and screening-oriented discovery. These benchmarks primarily assess predictive accuracy or screening quality, rather than the behavior of algorithms that actively generate candidates for specified targets.

\paragraph{Benchmarks for molecular and materials generation.}
Generative-model evaluation has been extensively studied in molecular design. QM9~\cite{ramakrishnan2014qm9} provides quantum-chemical molecular properties, MoleculeNet~\cite{wu2018moleculenet} supports molecular representation learning, and GuacaMol~\cite{brown2019guacamol} and MOSES~\cite{polykovskiy2020moses} evaluate de novo molecular generation, distribution learning, and goal-directed optimization. MGB~\cite{yan2025mgb} extends standardized evaluation to crystal structure prediction. These benchmarks mainly focus on structural validity, diversity, and distributional generalization, whereas formulation design requires evaluating target-conditioned optimization behavior under oracle and budget constraints.

\paragraph{Materials inverse design algorithms.}
Materials inverse design has been approached through several algorithmic families. Active learning and adaptive sampling have been used to reduce experimental or computational cost in materials discovery~\cite{xue2016accelerated,lookman2019active}. Bayesian optimization provides an uncertainty-aware framework for expensive black-box optimization~\cite{shahriari2016bayesian,frazier2018tutorial}, while genetic and surrogate-assisted search methods have been applied to explore complex materials design spaces~\cite{jennings2019genetic}. Deep generative models, including variational autoencoders~\cite{kingma2014auto} and generative adversarial networks~\cite{goodfellow2014generative}, enable latent or implicit candidate generation. Diffusion models further improve generative modeling through denoising and score-based sampling~\cite{ho2020denoising,song2021score,karras2022edm}.

\section{Inverse Benchmark Design}
\label{headings}

\begin{figure}[t]
    \centering
    \includegraphics[width=\linewidth]{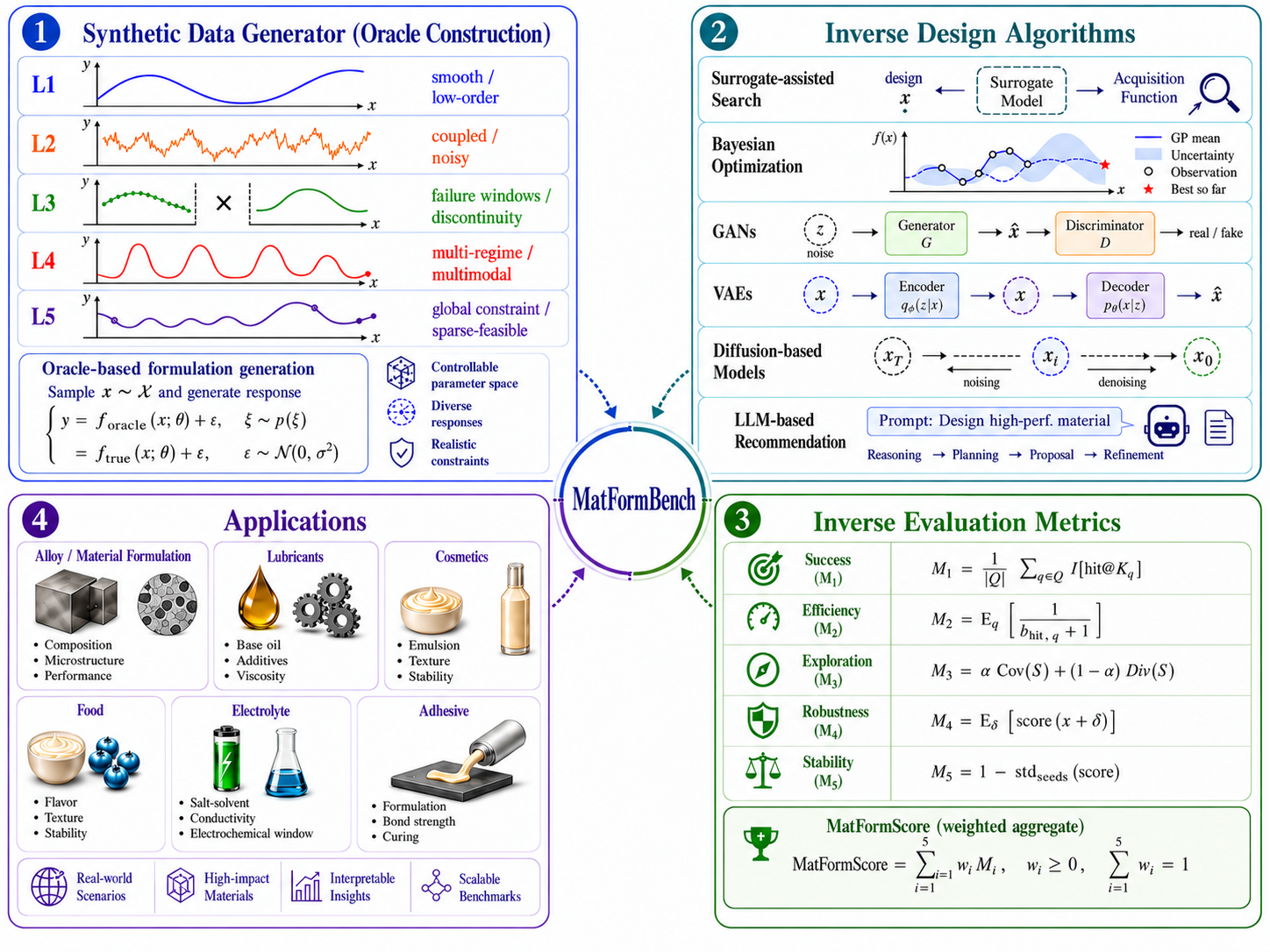}
    \caption{\textbf{Overview of MatFormBench.} MatFormBench integrates controllable synthetic oracle construction, heterogeneous inverse design algorithms, multi-axis inverse evaluation metrics, and representative formulation applications.}
    \label{fig:dvbench_framework}
\end{figure}

MatFormBench is designed as a unified evaluation framework for target-driven materials formulation. As shown in Figure~\ref{fig:dvbench_framework}, it combines controllable synthetic oracle construction, heterogeneous inverse design algorithms, and multi-axis evaluation metrics, while representative formulation scenarios illustrate its downstream relevance. The benchmark defines five difficulty levels, \(L1\)--\(L5\), to emulate increasingly complex materials response relationships, and evaluates algorithms using MatFormScore, which jointly measures success, efficiency, exploration, robustness, and stability.

\subsection{Problem Definition and Task Formulation}

Traditional materials machine learning is commonly formulated as forward property prediction. Let \(\mathcal{X}\subset\mathbb{R}^d\) denote the design space of compositions, structures, or processing conditions, and let \(\mathcal{Y}\subseteq\mathbb{R}^m\) denote the property space. The underlying property map is
\begin{equation}
f:\mathcal{X}\rightarrow\mathcal{Y}.
\end{equation}
Given observations \(\mathcal{D}=\{(x_i,y_i)\}_{i=1}^{n}\), with \(y_i=f(x_i)\), forward prediction learns an approximation \(\hat f_\theta\) such that \(\hat y=\hat f_\theta(x)\). This setting evaluates the mapping \(x\mapsto y\), i.e., property estimation for given material candidates.

In contrast, materials inverse design starts from a target property specification \(y^\star\in\mathcal{Y}\) and aims to identify or generate candidates \(x\in\mathcal{X}\) satisfying the desired target, corresponding to the reverse direction \(y^\star\mapsto x\). A basic formulation is
\begin{equation}
\min_{x\in\mathcal{X}} \ell(f(x),y^\star),
\end{equation}
where \(\ell:\mathcal{Y}\times\mathcal{Y}\rightarrow\mathbb{R}_{\ge 0}\) measures target mismatch. With additional feasibility, stability, or synthesizability requirements, inverse design becomes
\begin{equation}
\min_{x\in\mathcal{X}} \ell(f(x),y^\star)
\quad
\mathrm{s.t.}
\quad
g(x,f(x))\le 0,
\end{equation}
where \(g\) denotes a constraint map. Thus, inverse design can be viewed as a target-driven black-box constrained optimization or generation problem.

In practice, \(f\) is usually accessible only through experiments, simulations, or oracle evaluations, and must be queried under a finite budget \(B\). Given the history \(\mathcal{H}_{t-1}=\{(x_i,f(x_i))\}_{i=1}^{t-1}\), an algorithm selects the next candidate by
\begin{equation}
x_t\sim \pi_t(\cdot\mid\mathcal{H}_{t-1},y^\star),
\qquad t=1,\ldots,B.
\end{equation}
The central challenge is therefore not only accurate property prediction, but efficient target-conditioned candidate generation under limited oracle evaluations.

\subsection{Synthetic Data Construction with Difficulty Levels L1--L5}

MatFormBench constructs synthetic inverse-design tasks by progressively activating structural components in a controllable oracle. Difficulty is defined by both response-landscape complexity and feasible-region structure, following the hierarchy
\[
\text{smooth}
\rightarrow
\text{coupled/noisy}
\rightarrow
\text{locally discontinuous}
\rightarrow
\text{multimodal}
\rightarrow
\text{globally constrained}.
\]
Thus, \(L1\)--\(L5\) respectively capture smooth feasible responses, coupled noisy responses, local failures, multi-regime multimodality, and globally constrained sparse feasibility. Materials-inspired interpretations are provided in Appendix~\ref{app:materials_interpretation}.
\begin{table}[t]
\centering
\caption{Component activation pattern of the synthetic difficulty levels.}
\label{tab:difficulty_activation}
\small
\begin{tabular}{c|ccccc}
\toprule
Level & \(b(x)\) & \(c(x)\) & \(\Omega_{\mathrm{fail}}\) & \(m(x)\) & \(\phi(x)\leq\tau_s\) \\
\midrule
L1 & \(\checkmark\) & -- & -- & -- & -- \\
L2 & \(\checkmark\) & \(\checkmark\) & -- & -- & -- \\
L3 & \(\checkmark\) & \(\checkmark\) & \(\checkmark\) & -- & -- \\
L4 & \(\checkmark\) & \(\checkmark\) & \(\checkmark\) & \(\checkmark\) & -- \\
L5 & \(\checkmark\) & \(\checkmark\) & \(\checkmark\) & \(\checkmark\) & \(\checkmark\) \\
\bottomrule
\end{tabular}
\end{table}

Each design is represented as \(x=(x_1,\ldots,x_d)\in\mathcal{X}=[-1,1]^d\), with \(d\in\{5,10, 15\}\). At difficulty level \(L\), the oracle \(f_L:\mathcal{X}\rightarrow\mathbb{R}^3\cup\{\bot\}\) returns a valid property vector only when the feasibility indicator \(\psi_L(x)=1\):
\begin{equation}
f_L(x)=
\begin{cases}
(y_1(x),y_2(x),y_3(x)), & \psi_L(x)=1,\\
\bot, & \psi_L(x)=0.
\end{cases}
\end{equation}
Here, \(\bot\) denotes an invalid design. MatFormBench treats \(y_1\) as a maximization objective and \(y_2,y_3\) as minimization objectives. For feasible designs, the clean oracle is defined as
\begin{equation}
y_1^{\mathrm{clean}}(x)
=
60+\frac{20}{d}\bigl[b(x)+c_L(x)+m_L(x)\bigr]-0.01r_L(x).
\end{equation}

\begin{equation}
y_2^{\mathrm{clean}}(x)
=
200
\left(1+\frac{1}{d}\sum_{i=1}^{d}|x_i|\right)
\left(1+\frac{0.5}{100}\max\{r_L(x),0\}\right)s_L(x).
\end{equation}

\begin{equation}
y_3^{\mathrm{clean}}(x)
=
5
+
0.5\sum_{i=1}^{d}\max\{x_i,0\}
+
0.05\max\{r_L(x),0\}.
\end{equation}

The smooth backbone is \(b(x)=a(x)+q(x)+p(x)\), where \(a(x)=w^\top x\), \(q(x)=\frac{1}{2}\sum_i(\alpha_i x_i)^2\), and \(p(x)=0.3\sin(\pi\sum_{i=1}^{3}x_i)\). The coupling term \(c(x)\), multi-regime term \(m(x)\), correction terms \(r_L(x)\) and \(s_L(x)\), local failure set \(\Omega_{\mathrm{fail}}\), and global statistic \(\phi(x)\) are activated according to Table~\ref{tab:difficulty_activation}. The level-specific oracle definitions are summarized in Table~\ref{tab:l1_l5_formulas}.

\begin{table}[t]
\centering
\caption{Level-specific oracle definitions for the synthetic difficulty hierarchy.}
\label{tab:l1_l5_formulas}
\small
\renewcommand{\arraystretch}{1.25}
\resizebox{\linewidth}{!}{
\begin{tabular}{c|p{0.58\linewidth}|p{0.30\linewidth}}
\toprule
\multicolumn{1}{c|}{\textbf{Level}} &
\multicolumn{1}{c|}{\textbf{Definition}} &
\multicolumn{1}{c}{\textbf{Interpretation}} \\
\midrule

\textbf{\textit{L1}}
&
\(c_1=m_1=r_1=0,\ s_1=1,\ \psi_1(x)\equiv1\).
&
Smooth and globally feasible response.
\\

\midrule

\textbf{\textit{L2}}
&
\(c_2(x)=c(x),\ m_2=r_2=0,\ s_2=1,\ \psi_2(x)\equiv1\), with noisy observations \(\tilde y_j=y_j^{\mathrm{clean}}+\varepsilon_j(x)\).
&
Coupled and heteroscedastic response.
\\

\midrule

\textbf{\textit{L3}}
&
\(c_3(x)=c(x),\ m_3=0,\ \psi_3(x)=\mathbf{1}\{x\notin\Omega_{\mathrm{fail}}\}\), with rule corrections \(r_3,s_3\).
&
Local failure and discontinuity.
\\

\midrule

\textbf{\textit{L4}}
&
\(c_4(x)=c(x),\ m_4(x)=m(x),\ \psi_4(x)=\mathbf{1}\{x\notin\Omega_{\mathrm{fail}}\}\), with rule corrections \(r_4,s_4\).
&
Multi-regime multimodality.
\\

\midrule

\textbf{\textit{L5}}
&
\(c_5(x)=c(x),\ m_5(x)=m(x)\), and \(\psi_5(x)=\mathbf{1}\{x\notin\Omega_{\mathrm{fail}}\}\mathbf{1}\{\phi(x)\le\tau_s\}\).
&
Globally constrained sparse feasibility.
\\

\bottomrule
\end{tabular}
}
\end{table}

Training observations are generated from a noisy oracle \(\tilde f_L(x)=f_L^{\mathrm{clean}}(x)+\Delta_{\mathrm{batch}}(x)+\varepsilon_L(x)\), where \(\Delta_{\mathrm{batch}}\) denotes batch effects and \(\varepsilon_L\) denotes observational noise. Validation uses \(f_L^{\mathrm{clean}}\) with deterministic feasibility rules, ensuring that reported scores reflect the underlying inverse-design structure rather than random measurement fluctuations.

\subsection{Inverse Design Evaluation Metrics}

Inverse design evaluation differs from supervised prediction evaluation because the evaluated object is a sequential candidate-generation process under a finite oracle budget. A strong algorithm should not only generate target-satisfying candidates, but also find them early, explore the design space, remain effective under limited data, and produce stable results across random seeds. MatFormBench therefore evaluates algorithms along five axes: Success, Efficiency, Exploration, Robustness, and Stability.

Let an algorithm generate \(K\) candidates per round over \(T\) rounds, denoted by \(S^{(t)}=\{x_1^{(t)},\ldots,x_K^{(t)}\}\). For a task with \(M\) targets, define \(z_j(x)=\mathbf{1}\{x\text{ satisfies target }j\}\) and \(z(x)=\sum_{j=1}^{M}z_j(x)\). Thus, \(z(x)=M\) indicates all-target success, while invalid or infeasible outputs are treated as failures. The metric system is summarized in Table~\ref{tab:matformscore_metrics}.

\begin{table}[t]
\centering
\caption{Summary of MatFormBench inverse-design evaluation metrics.}
\label{tab:matformscore_metrics}
\small
\renewcommand{\arraystretch}{1.22}
\resizebox{\linewidth}{!}{
\begin{tabular}{c|c|c|c}
\toprule
\textbf{Axis} & \textbf{Statistic} & \textbf{Definition} & \textbf{Interpretation} \\
\midrule

\multirow{4}{*}{Success}
&
\(r^{(1)}_{\ge 1}\)
&
\(\frac{1}{K}\sum_{x\in S^{(1)}}\mathbf{1}\{z(x)\ge 1\}\)
&
First-round weak hit rate
\\
&
\(r^{(1)}_{\ge 2}\)
&
\(\frac{1}{K}\sum_{x\in S^{(1)}}\mathbf{1}\{z(x)\ge 2\}\)
&
First-round multi-target hit rate
\\
&
\(r^{(1)}_{\mathrm{all}}\)
&
\(\frac{1}{K}\sum_{x\in S^{(1)}}\mathbf{1}\{z(x)=M\}\)
&
First-round all-target hit rate
\\
&
\(S_{\mathrm{succ}}\)
&
\(0.20r^{(1)}_{\ge 1}+0.30r^{(1)}_{\ge 2}+0.50r^{(1)}_{\mathrm{all}}\)
&
Immediate target satisfaction
\\

\midrule

\multirow{4}{*}{Efficiency}
&
\(\tau_{\mathrm{first}}\)
&
\(\min\{t:\exists x\in S^{(t)},z(x)=M\}\)
&
First all-target success round
\\
&
\(E_{\mathrm{first}}\)
&
\(\exp[-(\tau_{\mathrm{first}}-1)/2.5]\)
&
Early-success reward
\\
&
\(E_{\mathrm{budget}}\)
&
\(0.20h^{(T)}_{\ge1}+0.30h^{(T)}_{\ge2}+0.50h^{(T)}_{\mathrm{all}}\)
&
Budgeted target satisfaction
\\
&
\(S_{\mathrm{eff}}\)
&
\(0.45E_{\mathrm{first}}+0.20\mathbf{1}\{N_{\mathrm{all}}\ge N_{\min}\}+0.35E_{\mathrm{budget}}\)
&
Early and sustained success
\\

\midrule

\multirow{3}{*}{Exploration}
&
\(D(X_K)\)
&
Normalized pairwise-distance diversity
&
Feature-space coverage
\\
&
\(\widetilde{\mathrm{HV}}\)
&
\(\mathrm{HV}(S_K)/(\mathrm{HV}(S_K)+1000)\)
&
Bounded Pareto hypervolume
\\
&
\(S_{\mathrm{exp}}\)
&
\(0.50D(X_K)+0.50\widetilde{\mathrm{HV}}\)
&
Diversity and frontier quality
\\

\midrule

\multirow{2}{*}{Robustness}
&
\(h_{\mathrm{all}}(n)\)
&
All-target hit rate using \(n\in\{10,15,30,50,100\}\) training samples
&
Data-scale performance
\\
&
\(S_{\mathrm{rob}}\)
&
\(\sum_{n\in\mathcal{N}}w_nh_{\mathrm{all}}(n)\)
&
Small-data weighted robustness
\\

\midrule

\multirow{3}{*}{Stability}
&
\(\mu_{\mathrm{seed}}\)
&
\(\frac{1}{|\mathcal{S}|}\sum_{s\in\mathcal{S}}h_s\)
&
Mean hit rate across seeds
\\
&
\(\sigma_{\mathrm{seed}}\)
&
\(\sqrt{\frac{1}{|\mathcal{S}|}\sum_{s\in\mathcal{S}}(h_s-\mu_{\mathrm{seed}})^2}\)
&
Seed-level variability
\\
&
\(S_{\mathrm{stab}}\)
&
\(\mu_{\mathrm{seed}}(1-\sigma_{\mathrm{seed}})\)
&
Seed-level reproducibility
\\

\bottomrule
\end{tabular}
}
\end{table}

For robustness, MatFormBench uses small-data-biased weights \((w_{10},w_{15},w_{30},w_{50},w_{100})=(0.35,0.25,0.20,0.12,0.08)\). The final MatFormScore aggregates the five normalized axes as
\begin{equation}
\mathrm{MatFormScore}
=
100\cdot
\mathrm{clip}_{[0,1]}
\left(
0.45S_{\mathrm{succ}}
+
0.25S_{\mathrm{eff}}
+
0.05S_{\mathrm{exp}}
+
0.15S_{\mathrm{rob}}
+
0.10S_{\mathrm{stab}}
\right).
\end{equation}
The weights prioritize strict target satisfaction and budget efficiency, while robustness, stability, and exploration provide diagnostic information about data sensitivity, reproducibility, and candidate diversity.

\subsection{Benchmark Protocol}

MatFormBench formulates benchmarking as a finite-budget sequential decision process. A task instance is \(\mathcal{I}=(\mathcal{X},f,y^\star,R,K)\), where \(\mathcal{X}\) is the design space, \(f\) is the oracle, \(y^\star\) is the target specification, \(R\) is the number of rounds, and \(K\) is the number of candidates submitted per round. At round \(t\), an algorithm \(\mathcal{A}\) submits \(S^{(t)}=\{x_1^{(t)},\ldots,x_K^{(t)}\}\subset\mathcal{X}\) according to \(S^{(t)}\sim\pi_t(\cdot\mid\mathcal{H}_{t-1},y^\star)\), where \(\mathcal{H}_{t-1}=\{(S^{(i)},f(S^{(i)}))\}_{i=1}^{t-1}\) is the observation history. The oracle evaluates \(S^{(t)}\), updates the history, and yields the trajectory \(\mathcal{P}=\{(S^{(t)},f(S^{(t)}))\}_{t=1}^{R}\).

All algorithms are evaluated on the same \((\mathcal{X},f,y^\star)\) and under the same total budget \(RK\). No method may access future observations or oracle information beyond \(\mathcal{H}_{t-1}\). Thus, learning-based, generative, Bayesian, and heuristic methods are compared through their effective candidate-selection behavior under matched resource constraints.

For each task, MatFormBench records the full trajectory and computes the metric components in Section~3.3. Across repeated dataset seeds, if \(D^{(m)}\) denotes the MatFormScore of the \(m\)-th repetition, the reported statistics are the empirical mean \(\bar D=\frac{1}{M}\sum_{m=1}^{M}D^{(m)}\) and variance \(\mathrm{Var}(D)=\frac{1}{M}\sum_{m=1}^{M}(D^{(m)}-\bar D)^2\). This protocol makes algorithmic comparisons interpretable as differences in finite-budget inverse-design policies under a common task, oracle, target, and evaluation functional.

\section{Experimental Setup}

\subsection{Benchmark Datasets}

Using the oracle family defined in Section~3.2, MatFormBench instantiates \(30\) benchmark datasets, denoted as \(\mathcal{D}_{\mathrm{bench}}=\{\mathcal{D}_{\ell,k}:\ell=1,\ldots,5,\ k=1,\ldots,6\}\). Each dataset is defined on \(\mathcal{X}=[-1,1]^d\) and corresponds to a three-objective target region \(\mathcal{T}_{\ell,k}=\{x\in\mathcal{X}:y_1(x)\ge\tau_1,\ y_2(x)\le\tau_2,\ y_3(x)\le\tau_3\}\). For \(L1\)--\(L4\), Datasets 1--3 use \(d=5\), while Datasets 4--6 use \(d=10\). For \(L5\), Datasets 1--3 use \(d=10\), and Datasets 4--6 use \(d=15\). Within each group, the thresholds are progressively tightened, making the target region increasingly extrapolative. Thus, \(\ell\) controls oracle-structural difficulty, while \(k\) controls dimensionality and target sparsity.

\subsection{Benchmark Inverse Design Algorithms}

We evaluate \(39\) inverse design algorithms that represent mainstream formulation-generation and inverse-optimization paradigms in AI for Science. Formally, the algorithm suite is \(\mathcal{A}_{\mathrm{bench}}=\mathcal{A}_{\mathrm{LLM}}\cup\mathcal{A}_{\mathrm{Diff}}\cup\mathcal{A}_{\mathrm{VAE}}\cup\mathcal{A}_{\mathrm{GAN}}\cup\mathcal{A}_{\mathrm{BO}}\cup\mathcal{A}_{\mathrm{search}}\), where each method induces a candidate-generation rule \(S\sim\pi_{\mathcal{A}}(\cdot\mid\mathcal{D}_{\mathrm{train}},y^\star)\). As summarized in Table~\ref{tab:benchmark_algorithms}, the benchmark includes an LLM-based direct recommendation baseline, diffusion models, VAE and GAN generative models, Bayesian optimization methods, and surrogate-assisted search algorithms. Classical search methods are paired with surrogate models \(\hat f_\theta\) because unevaluated candidates cannot be directly ranked by the clean oracle. Ridge regression, Gaussian process regression (GPR), and support vector regression (SVR) are used as surrogates due to their comparatively stable extrapolative behavior.

\begin{table}[t]
\centering
\caption{Benchmark inverse design algorithms evaluated in MatFormBench.}
\label{tab:benchmark_algorithms}
\scriptsize
\renewcommand{\arraystretch}{1.18}
\begin{tabularx}{\linewidth}{l X c}
\toprule
\multicolumn{1}{c}{\textbf{Family}} &
\multicolumn{1}{c}{\textbf{Methods}} &
\multicolumn{1}{c}{\textbf{Count}} \\
\midrule

LLM-based recommendation 
&
DeepSeek-v4-flash~\cite{deepseek2026v4}, 
GLM-5.1~\cite{glm2026technical}, 
Kimi K2.6 ~\cite{kimi2026technical}
& 3 \\

Diffusion models 
&
DDIM~\cite{song2020ddim}, 
DDPM~\cite{ho2020denoising}, 
EDM~\cite{karras2022edm}, 
Flow Matching~\cite{lipman2022flow}
& 4 \\

VAE models 
&
BetaVAE~\cite{higgins2017betavae}, 
CVAE~\cite{sohn2015cvae}, 
IWAE~\cite{burda2016iwae}, 
VampPrior~\cite{tomczak2018vamprior}
& 4 \\

GAN models 
&
CGAN~\cite{mirza2014cgan}, 
CTGAN~\cite{xu2019ctgan}, 
PacGAN~\cite{lin2018pacgan}, 
WGAN-GP~\cite{gulrajani2017wgangp}
& 4 \\

Bayesian optimization 
&
GoalPA~\cite{shahriari2016bayesian}, 
RGP-UCB~\cite{srinivas2010gpucb}, 
TargetEGO~\cite{jones1998ego}
& 3 \\


Surrogate-assisted search
&
GA, PSO, MCTS, SA, Wolf, ACO, and Firefly optimizers paired with Ridge, GPR, and SVR surrogates~\cite{holland1975ga,kennedy1995particle,browne2012mcts,kirkpatrick1983optimization,mirjalili2014gwo,dorigo1996ant,yang2009firefly,hoerl1970ridge,rasmussen2006gp,cortes1995svm}
& 21 \\

\midrule
Total & -- & 39 \\
\bottomrule
\end{tabularx}
\end{table}

\subsection{Training and Search Budget}

For each dataset, the default initial training set contains \(n_0=30\) Latin hypercube samples from \(\mathcal{X}=[-1,1]^d\). Each non-LLM method first generates a candidate pool of size \(N_{\mathrm{gen}}=100\) and submits the top \(K=5\) candidates for clean oracle validation; the LLM baselines directly submit \(K=5\) candidates. Closed-loop evaluation uses \(R=5\) rounds, yielding a total validation budget of \(B=RK=25\) candidates per algorithm-task pair. After each round, validated candidates are appended to the training history as \(\mathcal{D}_{t}=\mathcal{D}_{t-1}\cup\{(x,f^{\mathrm{clean}}(x)):x\in S^{(t)}\}\). For robustness analysis, the initial training size is varied over \(\mathcal{N}=\{10,15,30,50,100\}\), while \(K\), \(R\), and the target constraints remain fixed.

\subsection{Reproducibility and Fairness Settings}

All algorithms are evaluated on the same task tuple \((\mathcal{X},f_\ell,y^\star)\), target constraints, and clean validation oracle \(f_\ell^{\mathrm{clean}}\). Training sets are sampled using fixed dataset seeds and the same Latin hypercube design rule. Stability is evaluated over \(\mathcal{S}=\{11,22,33,44,55,66,77,88,99,111\}\). Invalid candidates, infeasible outputs, and missing values are treated as failures. For each run, MatFormBench stores submitted candidates, clean oracle values, hit statistics, diversity, hypervolume, robustness, stability, and final MatFormScore in a unified JSON record.

\section{Experimental Results and Analysis}

We evaluate MatFormBench on \(30\) benchmark datasets. In total, \(39\) algorithms are attempted; \(37\) algorithms produce valid oracle-evaluable outputs. The two failed cases, GLM-5.1~\cite{glm2026technical} and Kimi K2.6~\cite{kimi2026technical}, do not return valid candidate formulations under the required output protocol and are therefore treated as failed LLM trials. Unless otherwise specified, LLM-related scores are computed from the valid DeepSeek-v4-flash~\cite{deepseek2026v4} baseline only.

\subsection{Overall Benchmark Results}

\begin{figure}[t]
\centering

\begin{subfigure}[t]{0.49\linewidth}
    \centering
    \includegraphics[
        width=\linewidth,
        height=0.24\textheight,
        keepaspectratio
    ]{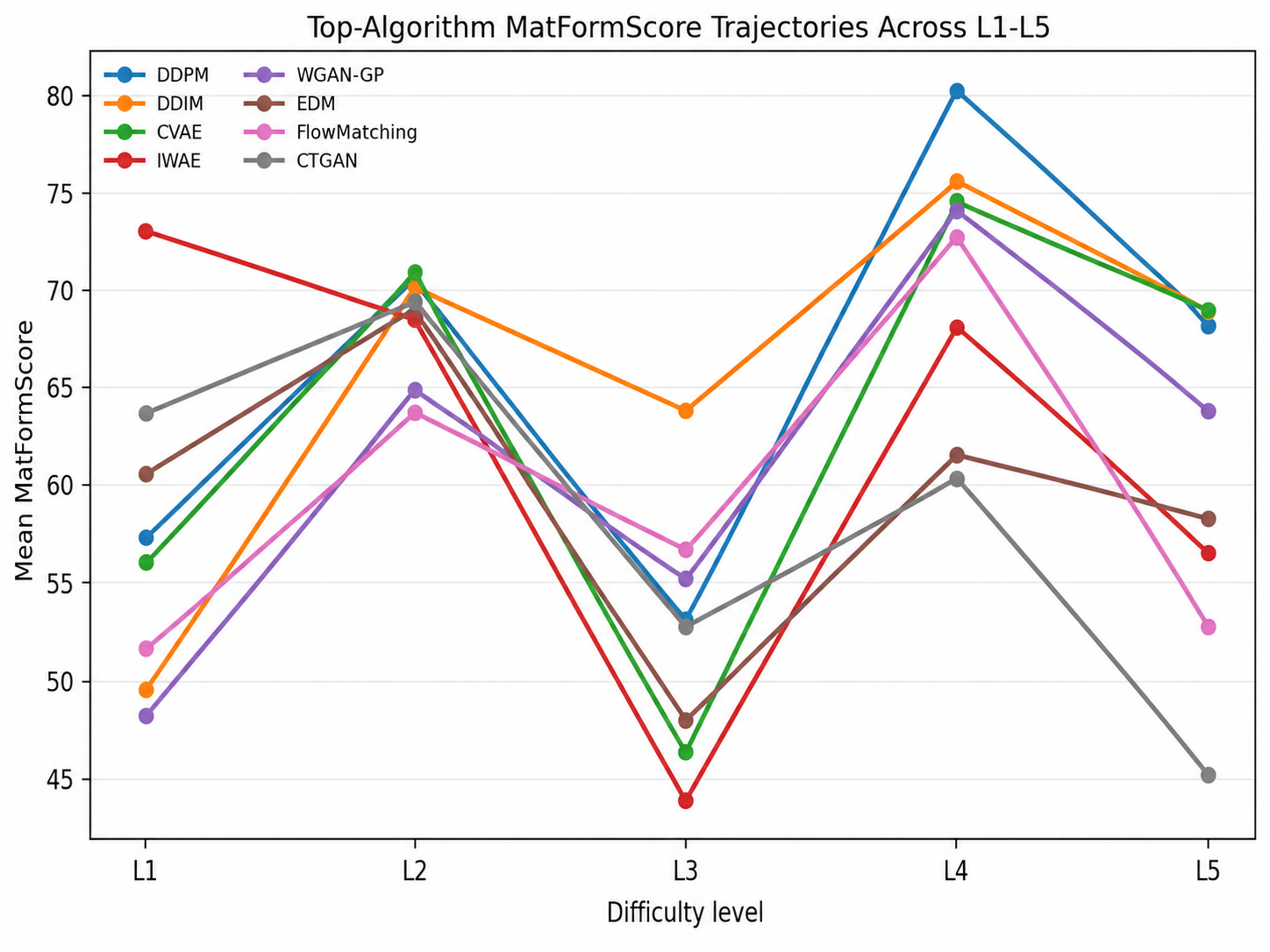}
    \caption{Top-algorithm MatFormScore trajectories across \(L1\)--\(L5\).}
    \label{fig:overall_results_a}
\end{subfigure}
\hfill
\begin{subfigure}[t]{0.49\linewidth}
    \centering
    \includegraphics[
        width=\linewidth,
        height=0.24\textheight,
        keepaspectratio
    ]{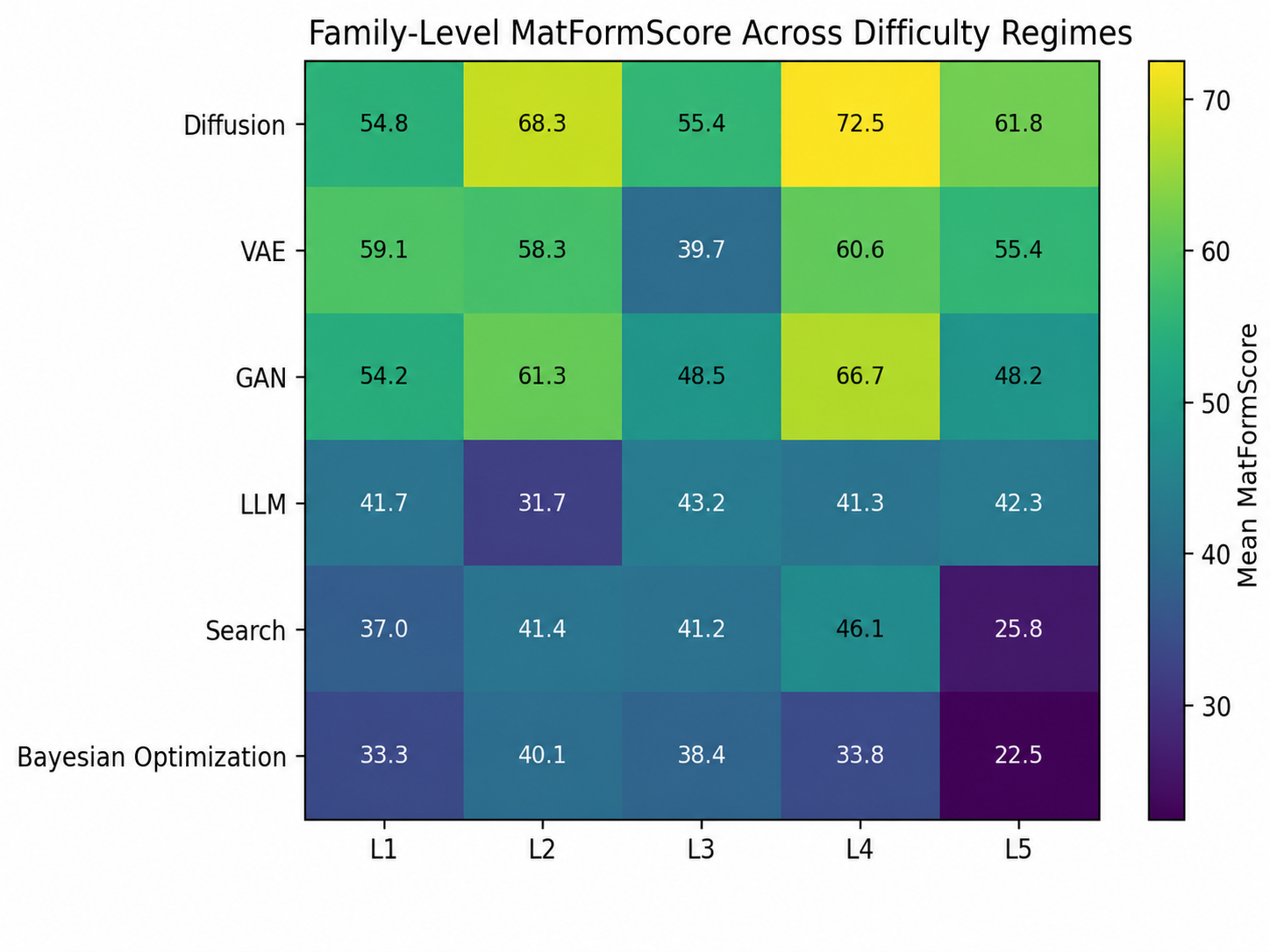}
    \caption{Family-level MatFormScore heatmap.}
    \label{fig:overall_results_b}
\end{subfigure}

\caption{\textbf{Overall benchmark performance.} Diffusion-based methods achieve the strongest aggregate performance and remain consistently competitive across difficulty regimes.}
\label{fig:overall_results}
\end{figure}

Figure~\ref{fig:overall_results} summarizes the overall performance. Diffusion models dominate the benchmark: DDPM~\cite{ho2020denoising} and DDIM~\cite{song2020ddim} achieve the highest mean MatFormScores, \(65.79\) and \(65.54\), respectively, followed by CVAE~\cite{sohn2015cvae}, IWAE~\cite{burda2016iwae}, and WGAN-GP~\cite{gulrajani2017wgangp}. Among classical methods, GA-SVR~\cite{holland1975ga,cortes1995svm} and GA-Ridge~\cite{holland1975ga,hoerl1970ridge} are the strongest baselines, indicating that surrogate-assisted evolutionary search remains competitive but still trails the leading diffusion models.

\begin{table}[t]
\centering
\caption{Top-performing algorithms ranked by mean MatFormScore.}
\label{tab:top_algorithms}
\small
\renewcommand{\arraystretch}{1.10}
\resizebox{\linewidth}{!}{
\begin{tabular}{c l l c c c c c}
\toprule
\textbf{Rank} & \textbf{Algorithm} & \textbf{Family} & \textbf{MatFormScore} & \textbf{Success} & \textbf{Efficiency} & \textbf{Exploration} & \textbf{Stability} \\
\midrule
1 & DDPM~\cite{ho2020denoising} & Diffusion & 65.79 & 0.677 & 0.833 & 0.816 & 0.368 \\
2 & DDIM~\cite{song2020ddim} & Diffusion & 65.54 & 0.664 & 0.816 & 0.820 & 0.386 \\
3 & CVAE~\cite{sohn2015cvae} & VAE & 63.29 & 0.645 & 0.835 & 0.791 & 0.350 \\
4 & IWAE~\cite{burda2016iwae} & VAE & 61.96 & 0.669 & 0.824 & 0.751 & 0.234 \\
5 & WGAN-GP~\cite{gulrajani2017wgangp} & GAN & 61.16 & 0.613 & 0.817 & 0.796 & 0.319 \\
6 & EDM~\cite{karras2022edm} & Diffusion & 59.49 & 0.610 & 0.758 & 0.804 & 0.301 \\
7 & Flow Matching~\cite{lipman2022flow} & Diffusion & 59.43 & 0.579 & 0.825 & 0.798 & 0.308 \\
8 & CTGAN~\cite{xu2019ctgan} & GAN & 58.27 & 0.584 & 0.807 & 0.765 & 0.257 \\
9 & GA-SVR~\cite{holland1975ga,cortes1995svm} & Search & 56.89 & 0.611 & 0.686 & 0.812 & 0.221 \\
10 & GA-Ridge~\cite{holland1975ga,hoerl1970ridge} & Search & 55.71 & 0.578 & 0.702 & 0.797 & 0.222 \\
\bottomrule
\end{tabular}
}
\end{table}

\subsection{Performance Across Task Regimes}

\begin{figure}[t]
\centering
\begin{subfigure}[t]{0.49\linewidth}
    \centering
    \begin{minipage}[t][0.23\textheight][c]{\linewidth}
        \centering
        \includegraphics[
            width=\linewidth,
            height=0.23\textheight,
            keepaspectratio
        ]{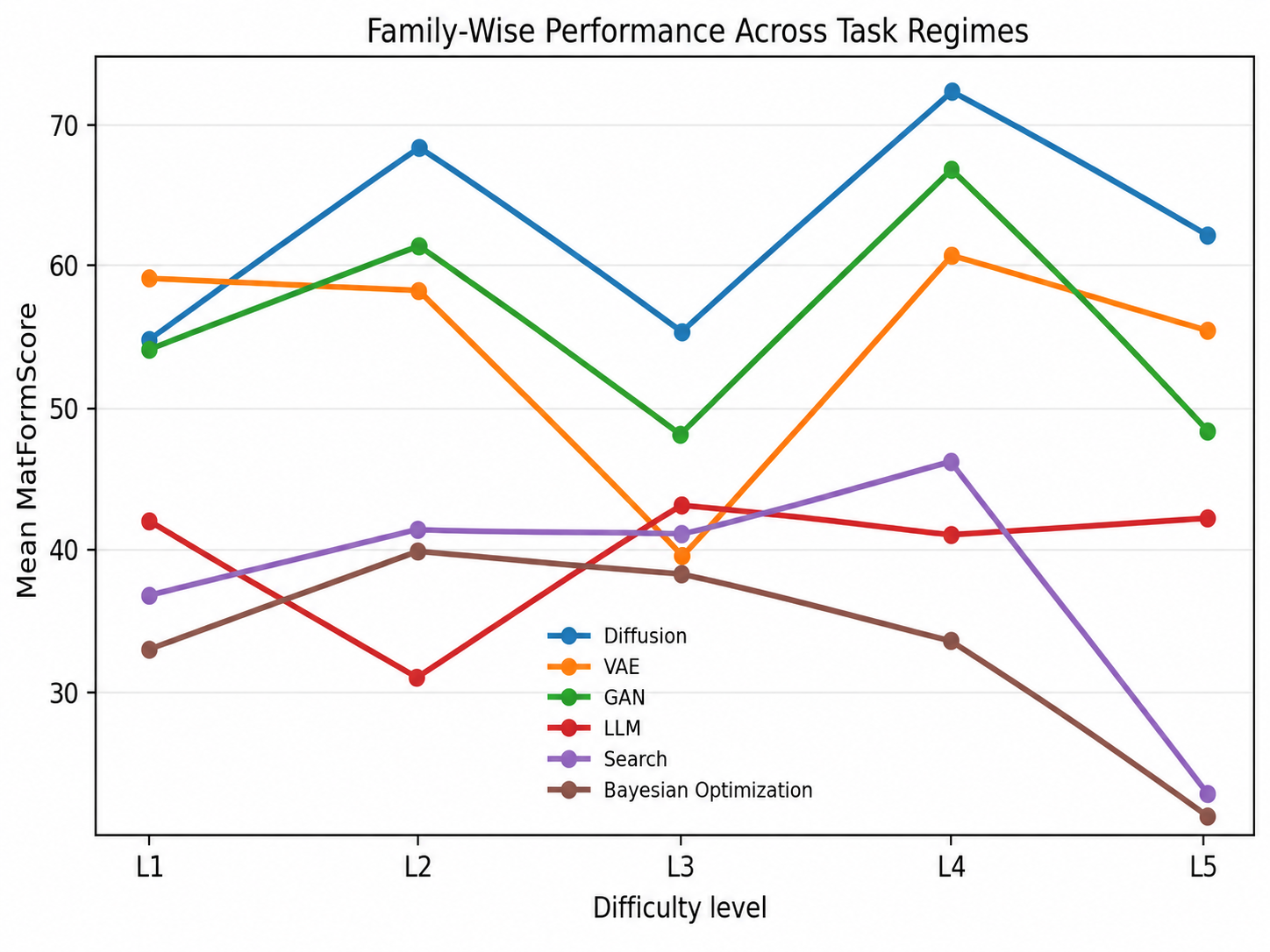}
    \end{minipage}
    \caption{Family-wise MatFormScore curves.}
    \label{fig:regime_performance_a}
\end{subfigure}
\hfill
\begin{subfigure}[t]{0.49\linewidth}
    \centering
    \begin{minipage}[t][0.23\textheight][c]{\linewidth}
        \centering
        \includegraphics[
            width=\linewidth,
            height=0.23\textheight,
            keepaspectratio
        ]{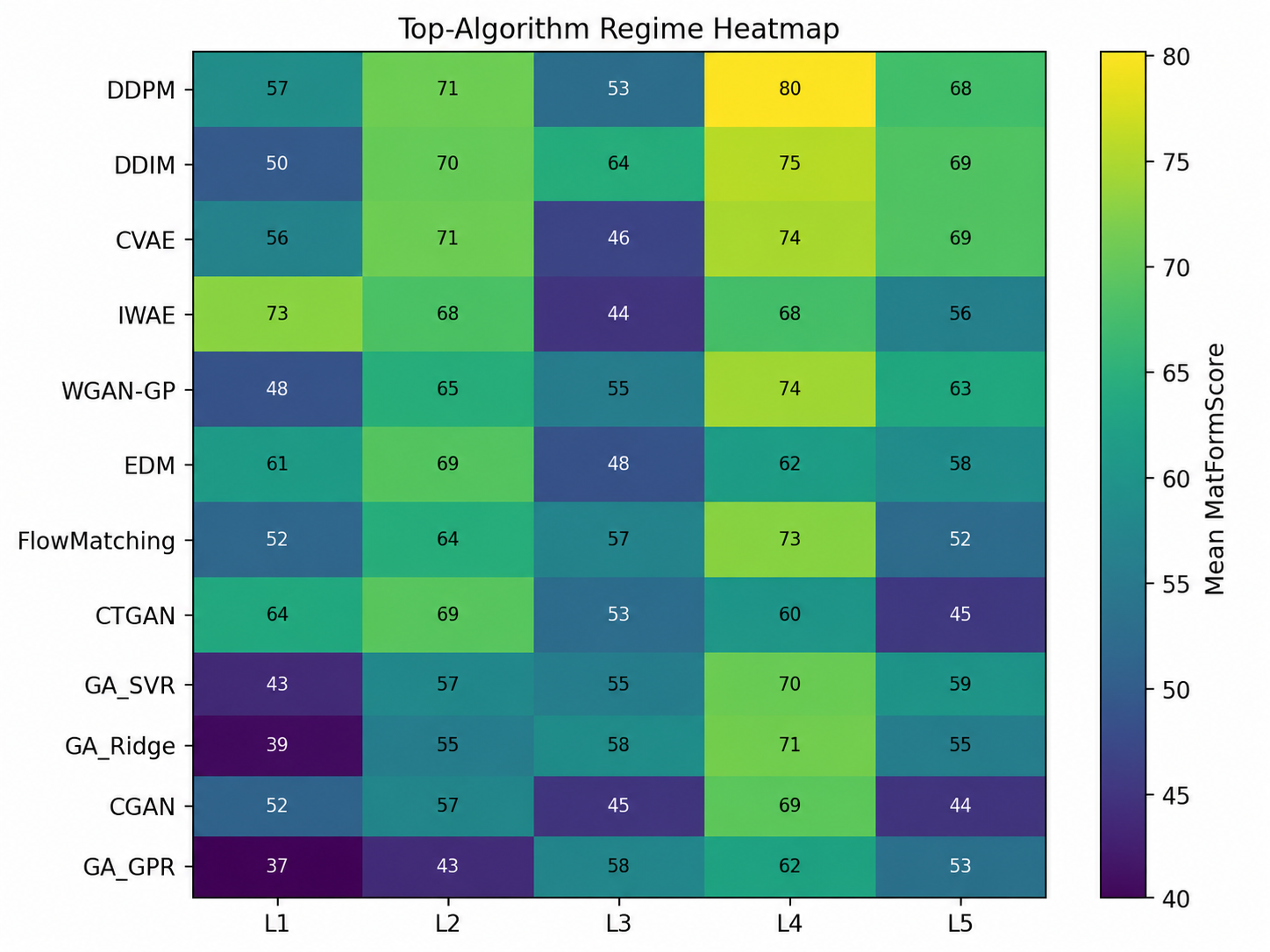}
    \end{minipage}
    \caption{Top-algorithm regime heatmap.}
    \label{fig:regime_performance_b}
\end{subfigure}
\caption{\textbf{Performance across task regimes.} MatFormBench reveals clear regime dependence: VAE methods are strong on smooth tasks, GA-based search is competitive under local discontinuity, and diffusion models dominate multimodal and globally constrained regimes.}
\label{fig:regime_performance}
\end{figure}

Figure~\ref{fig:regime_performance} shows that algorithm ranking varies across \(L1\)--\(L5\). VAE-based models perform best on the smooth \(L1\) tasks, where IWAE~\cite{burda2016iwae} reaches \(72.93\). On \(L2\), CVAE~\cite{sohn2015cvae}, DDPM~\cite{ho2020denoising}, and DDIM~\cite{song2020ddim} achieve comparable scores, indicating that conditional generative models handle coupled and noisy responses effectively. Under \(L3\), GA-based surrogate search~\cite{holland1975ga} becomes competitive, suggesting that local discontinuities weaken purely generative sampling. Diffusion models~\cite{ho2020denoising,song2021score,karras2022edm} dominate the multimodal \(L4\) regime, with DDPM reaching \(80.19\), and remain highly competitive under the sparse-feasible \(L5\) regime.

Overall, diffusion is the most consistent family across the difficulty hierarchy. Although VAE variants win some individual levels, diffusion achieves the strongest family-level performance over most regimes, especially under multimodality and global sparse feasibility.

\subsection{Algorithm Suitability Analysis}

\begin{figure}[t]
\centering
\begin{subfigure}[t]{0.49\linewidth}
    \centering
    \vspace{0pt}
    \includegraphics[height=0.28\textheight, keepaspectratio]{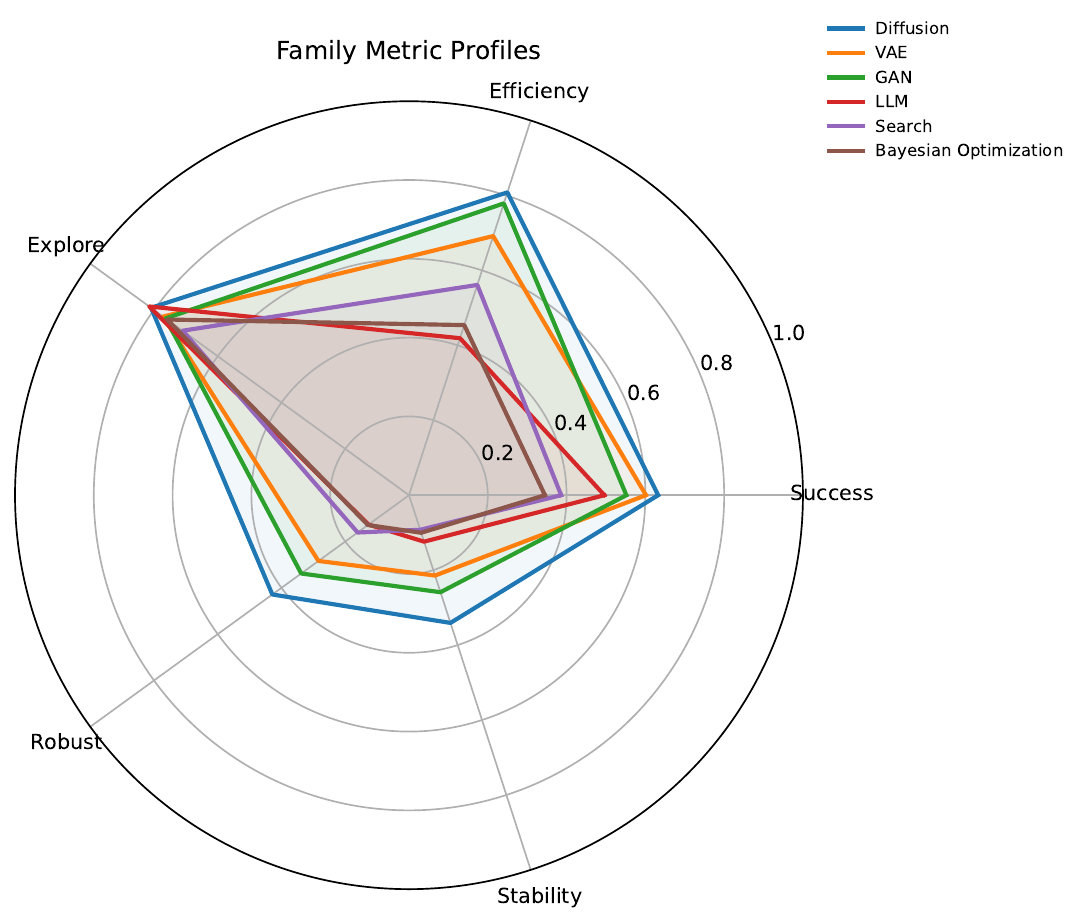}
    \caption{All algorithm families.}
    \label{fig:suitability_analysis_a}
\end{subfigure}
\hfill
\begin{subfigure}[t]{0.49\linewidth}
    \centering
    \vspace{0pt}
    \includegraphics[height=0.28\textheight, keepaspectratio]{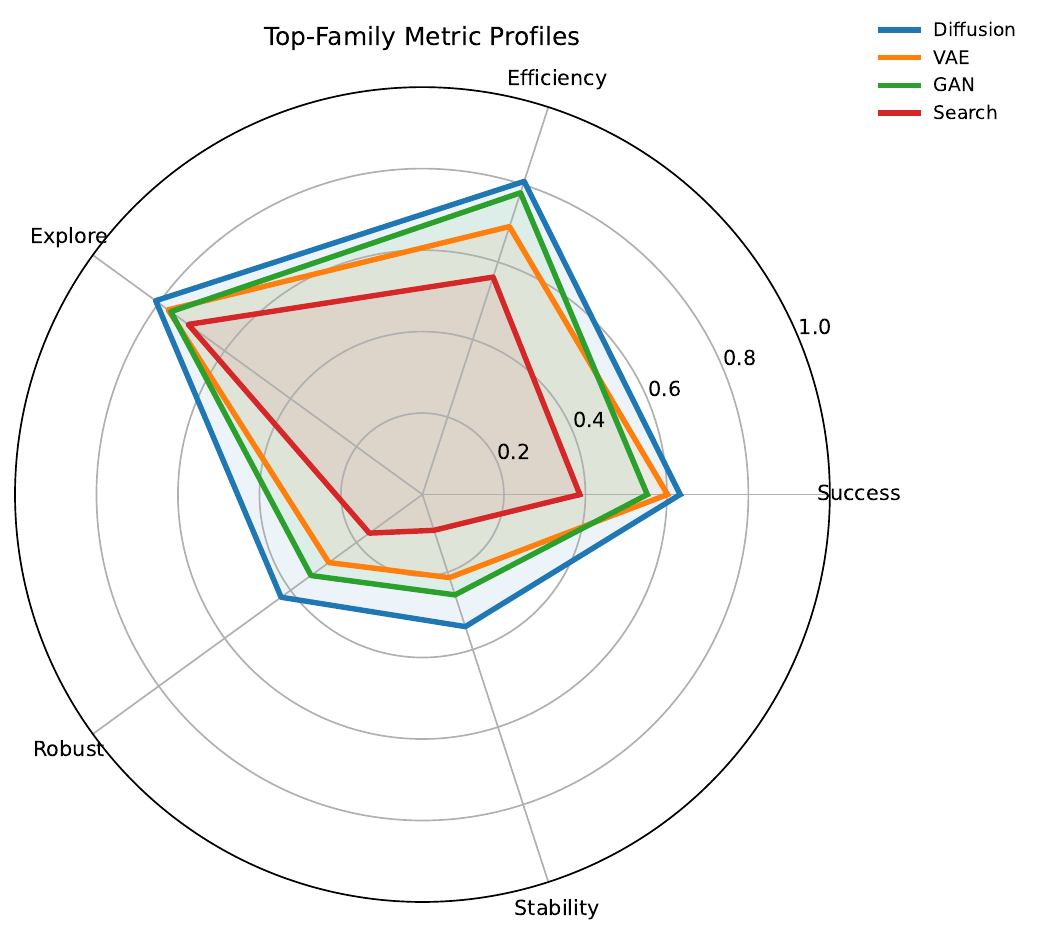}
    \caption{Top-performing families.}
    \label{fig:suitability_analysis_b}
\end{subfigure}
\caption{\textbf{Algorithm suitability analysis.} Radar plots compare family-level profiles over Success, Efficiency, Explore, Robustness, and Stability.}
\label{fig:suitability_analysis}
\end{figure}

We further compare algorithm families along the five MatFormScore axes: Success, Efficiency, Explore, Robustness, and Stability. As shown in Figure~\ref{fig:suitability_analysis} and Table~\ref{tab:family_metrics}, diffusion models~\cite{ho2020denoising,song2021score,karras2022edm} exhibit the most balanced profile, combining high success, efficiency, exploration, robustness, and stability. GAN~\cite{goodfellow2014generative} and VAE~\cite{kingma2014auto} methods form the second tier, while GA-based surrogate search~\cite{holland1975ga,jennings2019genetic} remains useful in discontinuous regimes.

\begin{table}[t]
\centering
\caption{Family-level metric profile. The LLM row is computed from the valid DeepSeek baseline only; GLM-5.1 and KIMI-2.6 fail to produce valid candidate outputs under the benchmark protocol.}
\label{tab:family_metrics}
\small
\renewcommand{\arraystretch}{1.10}
\resizebox{\linewidth}{!}{
\begin{tabular}{l c c c c c c}
\toprule
\textbf{Family} & \textbf{MatFormScore}& \textbf{Success} & \textbf{Efficiency}& \textbf{Exploration}& \textbf{Robust} & \textbf{Stability} \\
\midrule
Diffusion & 62.56 & 0.633 & 0.808 & 0.809 & 0.429 & 0.341 \\
GAN & 55.78 & 0.552 & 0.779 & 0.764 & 0.338 & 0.259 \\
VAE & 54.62 & 0.602 & 0.692 & 0.771 & 0.284 & 0.214 \\
LLM & 40.04 & 0.496 & 0.419 & 0.815 & 0.127 & 0.124 \\
Search & 38.30 & 0.386 & 0.561 & 0.710 & 0.161 & 0.092 \\
Bayesian Optimization & 33.61 & 0.345 & 0.454 & 0.759 & 0.129 & 0.100 \\
\bottomrule
\end{tabular}
}
\end{table}

The LLM results require separate interpretation. DeepSeek-v4-flash~\cite{deepseek2026v4} produces diverse candidates but shows substantially lower efficiency, robustness, and stability than diffusion-based models. GLM-5.1~\cite{glm2026technical} and Kimi K2.6~\cite{kimi2026technical} fail to generate valid structured outputs for oracle evaluation, suggesting that language-model-only recommendation is not yet reliable for strict continuous formulation inverse design without numerical optimization, constraint handling, or tool-based validation.

More experimental results and implementation details are provided in Appendix~\ref{app:supp_results}.

\section{Conclusion}

In this work, we propose MatFormBench, a  benchmarking framework for target-driven materials formulation. By combining controllable oracles, five-level task difficulty, a closed-loop evaluation protocol, and the multi-axis MatFormScore, MatFormBench enables reproducible and diagnostic comparison of heterogeneous inverse design algorithms. Across \(30\) benchmark datasets, \(39\) algorithms were attempted and \(37\) produced valid oracle-evaluable outputs. The results show that diffusion-based models achieve the strongest overall performance, while VAE- and GA-based methods exhibit advantages in smooth and locally discontinuous regimes, respectively. The LLM results further suggest that language-model-only recommendation remains limited for strict continuous formulation optimization without stronger numerical constraint handling.  Future work will extend the benchmark with more realistic datasets, richer design spaces, synthesis-aware constraints, and metrics for uncertainty, cost, safety, and experimental feasibility.

\bibliographystyle{plainnat}
\bibliography{references}


\appendix
\numberwithin{equation}{section}
\numberwithin{table}{section}
\numberwithin{figure}{section}

\section{Benchmark Dataset and Oracle Details}
\label{app:dataset_oracle}

\subsection{Oracle implementation details}

MatFormBench represents each candidate formulation as a bounded continuous vector 
\(x=(x_1,\ldots,x_d)\in[-1,1]^d\), with \(d\in\{5,10,15\}\). 
Beyond the oracle components summarized in the main text, the implementation includes several 
additional rules to make the synthetic tasks closer to practical formulation optimization. First, 
a subset of composition-like variables can be projected onto a simplex group. In the released 
implementation, when \(d\geq 6\), the first six coordinates may be normalized as a simplex group,
\begin{equation}
x_{\mathcal{G}}
\leftarrow
\frac{\max(x_{\mathcal{G}},\epsilon)}
{\sum_{i\in\mathcal{G}}\max(x_i,\epsilon)},
\qquad 
\mathcal{G}=\{1,\ldots,6\}.
\label{eq:app_simplex_projection}
\end{equation}
where \(\epsilon=10^{-2}\) is a minimum component value. This operation provides a simplified
continuous analogue of normalized composition fractions.

Second, local failure rules are attached to physically interpretable coordinates. The precipitation-like
failure window is assigned to the temperature-like coordinate \(x_3\), while the gelation-like failure
rule is assigned to the catalyst-like coordinate \(x_5\). The implementation uses
\begin{equation}
\Omega_{\mathrm{precip}}
=
\{x:-0.05\leq x_3\leq 0.10\},
\qquad
\Omega_{\mathrm{gel}}
=
\{x:x_5>0.60\}.
\label{eq:app_local_failure}
\end{equation}
For \(L3\)--\(L5\), candidates satisfying either rule are marked infeasible and receive no valid
property vector. For \(L5\), an additional global sparse-feasibility constraint is imposed:
\begin{equation}
\sum_{i\in \mathrm{Top3}(x)}x_i \leq 2.0,
\label{eq:app_global_constraint}
\end{equation}
where \(\mathrm{Top3}(x)\) denotes the indices of the three largest coordinates of \(x\). This rule
creates sparse feasible regions and prevents algorithms from trivially increasing several dominant
components simultaneously.

Third, the noisy training oracle includes batch-dependent response shifts. Let \(b_{\mathrm{batch}}\)
denote a batch-specific additive shift for \(y_1\), and \(s_{\mathrm{batch}}\) denote a multiplicative
batch factor for \(y_2\). The implementation samples from five batches, with
\begin{equation}
b_{\mathrm{batch}}\in\{0.0,0.6,-0.4,0.2,-0.2\},
\qquad
s_{\mathrm{batch}}\in\{1.00,1.12,1.18,0.92,1.06\}.
\label{eq:app_batch_effects}
\end{equation}
The validation oracle removes these stochastic batch effects and evaluates all submitted candidates
using the deterministic clean response. Thus, training data mimic noisy experimental observations,
whereas benchmark scores measure the underlying inverse-design capability.

\subsection{Task registry and target constraints}

MatFormBench contains \(30\) benchmark datasets, organized as \(5\) difficulty levels and \(6\)
datasets per level. The main text specifies the overall construction rule; Table~\ref{tab:app_dataset_registry}
provides the full task registry, including dimensionality and target thresholds. Each task is defined by
\begin{equation}
\mathcal{T}_{\ell,k}
=
\{x\in\mathcal{X}:y_1(x)\geq \tau_1,\; y_2(x)\leq \tau_2,\; y_3(x)\leq \tau_3\}.
\label{eq:app_target_region}
\end{equation}
The dataset index \(k\) controls target tightness and dimensionality, while the level \(\ell\) controls
the structural complexity of the oracle.

\begin{longtable}{c c c c c c}
\caption{Full MatFormBench task registry. Each task is defined by dimensionality \(d\) and three target constraints.}
\label{tab:app_dataset_registry}\\
\toprule
\textbf{Level} & \textbf{Dataset} & \(\boldsymbol{d}\) & \(\boldsymbol{y_1}\) target & \(\boldsymbol{y_2}\) target & \(\boldsymbol{y_3}\) target \\
\midrule
\endfirsthead
\toprule
\textbf{Level} & \textbf{Dataset} & \(\boldsymbol{d}\) & \(\boldsymbol{y_1}\) target & \(\boldsymbol{y_2}\) target & \(\boldsymbol{y_3}\) target \\
\midrule
\endhead
L1 & Dataset-1 & 5  & \(\geq 61\)   & \(\leq 315\) & \(\leq 6.0\) \\
L1 & Dataset-2 & 5  & \(\geq 65\)   & \(\leq 310\) & \(\leq 6.0\) \\
L1 & Dataset-3 & 5  & \(\geq 67\)   & \(\leq 288\) & \(\leq 6.0\) \\
L1 & Dataset-4 & 10 & \(\geq 63\)   & \(\leq 300\) & \(\leq 6.0\) \\
L1 & Dataset-5 & 10 & \(\geq 63\)   & \(\leq 260\) & \(\leq 6.0\) \\
L1 & Dataset-6 & 10 & \(\geq 64\)   & \(\leq 250\) & \(\leq 6.0\) \\
\midrule
L2 & Dataset-1 & 5  & \(\geq 63\)   & \(\leq 315\) & \(\leq 6.0\) \\
L2 & Dataset-2 & 5  & \(\geq 65\)   & \(\leq 300\) & \(\leq 6.0\) \\
L2 & Dataset-3 & 5  & \(\geq 66\)   & \(\leq 290\) & \(\leq 5.9\) \\
L2 & Dataset-4 & 10 & \(\geq 65\)   & \(\leq 280\) & \(\leq 6.5\) \\
L2 & Dataset-5 & 10 & \(\geq 67\)   & \(\leq 270\) & \(\leq 6.5\) \\
L2 & Dataset-6 & 10 & \(\geq 67.5\) & \(\leq 265\) & \(\leq 6.3\) \\
\midrule
L3 & Dataset-1 & 5  & \(\geq 59\) & \(\leq 305\) & \(\leq 6.0\) \\
L3 & Dataset-2 & 5  & \(\geq 61\) & \(\leq 295\) & \(\leq 6.0\) \\
L3 & Dataset-3 & 5  & \(\geq 63\) & \(\leq 285\) & \(\leq 5.8\) \\
L3 & Dataset-4 & 10 & \(\geq 63\) & \(\leq 305\) & \(\leq 6.3\) \\
L3 & Dataset-5 & 10 & \(\geq 64\) & \(\leq 285\) & \(\leq 6.3\) \\
L3 & Dataset-6 & 10 & \(\geq 67\) & \(\leq 280\) & \(\leq 6.3\) \\
\midrule
L4 & Dataset-1 & 10 & \(\geq 64\)   & \(\leq 305\) & \(\leq 6.3\) \\
L4 & Dataset-2 & 10 & \(\geq 65\)   & \(\leq 290\) & \(\leq 6.3\) \\
L4 & Dataset-3 & 10 & \(\geq 67\)   & \(\leq 288\) & \(\leq 6.3\) \\
L4 & Dataset-4 & 15 & \(\geq 63\)   & \(\leq 310\) & \(\leq 7.0\) \\
L4 & Dataset-5 & 15 & \(\geq 65\)   & \(\leq 300\) & \(\leq 7.0\) \\
L4 & Dataset-6 & 15 & \(\geq 66.5\) & \(\leq 298\) & \(\leq 7.0\) \\
\midrule
L5 & Dataset-1 & 10 & \(\geq 64\) & \(\leq 300\) & \(\leq 6.0\) \\
L5 & Dataset-2 & 10 & \(\geq 65\) & \(\leq 298\) & \(\leq 6.0\) \\
L5 & Dataset-3 & 10 & \(\geq 66\) & \(\leq 296\) & \(\leq 6.0\) \\
L5 & Dataset-4 & 15 & \(\geq 62\) & \(\leq 305\) & \(\leq 6.8\) \\
L5 & Dataset-5 & 15 & \(\geq 63\) & \(\leq 300\) & \(\leq 6.6\) \\
L5 & Dataset-6 & 15 & \(\geq 64\) & \(\leq 298\) & \(\leq 6.5\) \\
\bottomrule
\end{longtable}

\subsection{Dataset generation and filtering}

For each task, training candidates are sampled using Latin hypercube sampling in the design box.
To avoid trivial inverse-design instances, generated training rows that already satisfy all target
constraints are filtered before constructing the initial training set. For \(L3\)--\(L5\), the generator also
controls the fraction of infeasible rows so that the initial training set contains sufficient feasible
observations. In the implementation, the feasible fraction is encouraged to be at least \(70\%\) whenever
possible. This choice avoids degenerate training sets in which all observations are invalid, while still
preserving local and global feasibility constraints.

\subsection{Materials interpretation of difficulty levels}
\label{app:materials_interpretation}

Although MatFormBench uses controllable synthetic oracles rather than real experimental datasets,
the five difficulty levels are designed to abstract common response motifs in materials formulation
inverse design. Therefore, \(L1\)--\(L5\) should not be interpreted as five specific material systems,
but as five representative classes of formulation-response landscapes. This connection helps clarify
why the benchmark difficulty hierarchy is relevant to practical target-driven formulation optimization.

\paragraph{L1: Smooth mixture-property interpolation.}
The \(L1\) regime corresponds to materials systems in which properties vary smoothly with component
fractions and most candidate formulations remain feasible. A typical example is the rule of mixtures,
where an effective property \(P_{\mathrm{mix}}\) is approximated by a composition-weighted average:
\begin{equation}
P_{\mathrm{mix}}
=
\sum_{i=1}^{n} c_i P_i,
\qquad
\sum_{i=1}^{n}c_i=1,
\qquad
c_i\geq 0.
\label{eq:app_rule_of_mixtures}
\end{equation}
For elastic properties of composite or multiphase materials, the Voigt and Reuss bounds provide
smooth upper and lower estimates:
\begin{equation}
E_{\mathrm{V}}
=
\sum_{i=1}^{n}c_iE_i,
\qquad
\frac{1}{E_{\mathrm{R}}}
=
\sum_{i=1}^{n}\frac{c_i}{E_i}.
\label{eq:app_voigt_reuss_bounds}
\end{equation}
These relations motivate \(L1\), where inverse design mainly requires interpolation over a broad
feasible region. Representative tasks include polymer blend tuning, coating formulation, single-phase
alloy interpolation, and composite modulus optimization.

\paragraph{L2: Coupled transport or kinetic responses.}
The \(L2\) regime corresponds to formulation problems where responses remain broadly feasible but
depend on nonlinear interactions among composition, temperature, transport, and kinetic variables.
For example, ionic or electronic conductivity is often described by an Arrhenius-type relation,
\begin{equation}
\sigma(T,c)
=
\sigma_0(c)
\exp\left(
-\frac{E_a(c)}{k_{\mathrm{B}}T}
\right),
\label{eq:app_arrhenius_conductivity}
\end{equation}
where both the prefactor \(\sigma_0(c)\) and activation energy \(E_a(c)\) may depend on formulation
composition \(c\). In electrolyte formulation, conductivity can also be related to ion diffusivity and
concentration through a Nernst--Einstein-type expression:
\begin{equation}
\sigma
=
\frac{F^2}{RT}
\sum_i z_i^2 D_i c_i.
\label{eq:app_nernst_einstein}
\end{equation}
These relations motivate \(L2\), where component coupling and noisy observations make inverse
optimization harder than smooth mixture-property interpolation. Representative systems include
battery electrolytes, lubricants, polymer-additive formulations, and cement or concrete admixtures.

\paragraph{L3: Local failure and discontinuity.}
The \(L3\) regime corresponds to formulation tasks with local invalid regions caused by phase
separation, precipitation, gelation, curing failure, or loss of processability. A common thermodynamic
motif is the Flory--Huggins mixing free energy for a binary polymer mixture:
\begin{equation}
\frac{\Delta G_{\mathrm{mix}}}{RT}
=
\frac{\phi_1}{N_1}\ln\phi_1
+
\frac{\phi_2}{N_2}\ln\phi_2
+
\chi\phi_1\phi_2,
\label{eq:app_flory_huggins}
\end{equation}
where \(\phi_i\) are volume fractions, \(N_i\) are degrees of polymerization, and \(\chi\) is an
interaction parameter. Local phase stability can be characterized by the curvature of the mixing free
energy:
\begin{equation}
\frac{\partial^2 \Delta G_{\mathrm{mix}}}{\partial \phi^2}>0
\quad \mathrm{stable},
\qquad
\frac{\partial^2 \Delta G_{\mathrm{mix}}}{\partial \phi^2}<0
\quad \mathrm{unstable}.
\label{eq:app_spinodal_stability}
\end{equation}
Similarly, gelation can be approximated by a percolation-like threshold:
\begin{equation}
p(f-1)>1,
\label{eq:app_gelation_threshold}
\end{equation}
where \(p\) denotes the extent of reaction and \(f\) denotes functionality. These stability and gelation
criteria motivate \(L3\), where nearby formulations may switch abruptly from feasible to invalid.
Representative systems include emulsions, hydrogels, adhesives, curing resins, and ceramic slurries.

\paragraph{L4: Multi-regime and multimodal response mechanisms.}
The \(L4\) regime corresponds to materials systems in which multiple phases, microstructures, or
mechanisms can produce similar target properties. In phase-forming materials, the equilibrium state
is often governed by free-energy minimization:
\begin{equation}
G_{\mathrm{eq}}(x,T)
=
\min_{p\in\mathcal{P}}G_p(x,T),
\label{eq:app_phase_selection}
\end{equation}
where \(G_p(x,T)\) is the Gibbs free energy of phase or regime \(p\). A softened multi-regime
response can be written as a free-energy-weighted mixture:
\begin{equation}
P(x)
=
\sum_{r=1}^{R}\omega_r(x)P_r(x),
\qquad
\omega_r(x)
=
\frac{
\exp[-G_r(x)/(k_{\mathrm{B}}T)]
}{
\sum_{s=1}^{R}\exp[-G_s(x)/(k_{\mathrm{B}}T)]
}.
\label{eq:app_multiregime_response}
\end{equation}
These relations motivate \(L4\), where the inverse problem becomes multimodal because distinct
formulation families may satisfy the same target through different mechanisms. Representative
systems include high-entropy alloys, perovskite compositions, multi-additive polymer composites,
catalysts, and multiphase functional materials.

\paragraph{L5: Sparse feasibility under global constraints.}
The \(L5\) regime corresponds to high-dimensional formulation problems where target satisfaction
must be achieved together with global constraints such as cost, safety, density, toxicity, sustainability,
or processability. A generic constrained formulation objective can be written as
\begin{equation}
\min_{x\in\mathcal{X}}
\ell(f(x),y^\star)
\quad
\mathrm{s.t.}
\quad
C_{\mathrm{cost}}(x)\leq C_{\max},
\quad
R_{\mathrm{risk}}(x)\leq R_{\max},
\quad
\rho(x)\leq \rho_{\max}.
\label{eq:app_global_constraints}
\end{equation}
Composition-dependent cost and risk can be approximated by mixture rules:
\begin{equation}
C_{\mathrm{cost}}(x)
=
\sum_{i=1}^{n}c_iC_i,
\qquad
R_{\mathrm{risk}}(x)
=
\sum_{i=1}^{n}c_ir_i.
\label{eq:app_cost_risk_constraints}
\end{equation}
For lightweight structural materials, a specific-property constraint may also be imposed, for example
\begin{equation}
\frac{\sigma_y(x)}{\rho(x)}
\geq
\Gamma_{\min},
\label{eq:app_specific_strength}
\end{equation}
where \(\sigma_y(x)\) is yield strength, \(\rho(x)\) is density, and \(\Gamma_{\min}\) is a minimum
specific-strength requirement. These constraints motivate \(L5\), where high-performing candidates
are rare because they must satisfy both target properties and global feasibility requirements.
Representative systems include sustainable polymers, aerospace alloys, solid-state electrolytes,
battery electrolytes under safety constraints, and biomedical hydrogels or adhesives.

\begin{table}[t]
\centering
\caption{Materials interpretation of the MatFormBench difficulty hierarchy. The formulas are not used as exact oracle definitions, but serve as materials-inspired response motifs that motivate the five synthetic difficulty regimes.}
\label{tab:app_materials_interpretation}
\small
\renewcommand{\arraystretch}{1.18}
\resizebox{\linewidth}{!}{
\begin{tabular}{l l l l}
\toprule
\textbf{Level} & \textbf{Materials response type} & \textbf{Representative formula motif} & \textbf{Example systems} \\
\midrule
L1 & Smooth mixture-property interpolation 
& Rule of mixtures; Voigt--Reuss bounds 
& Polymer blends, coatings, single-phase alloys, composites \\
L2 & Coupled transport or kinetic response 
& Arrhenius conductivity; Nernst--Einstein relation 
& Electrolytes, lubricants, polymer additives, concrete admixtures \\
L3 & Local failure and discontinuity 
& Flory--Huggins free energy; spinodal stability; gelation threshold 
& Emulsions, hydrogels, adhesives, curing resins, ceramic slurries \\
L4 & Multi-regime and multimodal mechanisms 
& Gibbs free-energy phase selection; multi-regime response mixture 
& High-entropy alloys, perovskites, catalysts, polymer composites \\
L5 & Sparse feasibility under global constraints 
& Constrained optimization; cost/risk/density/specific-property constraints 
& Sustainable polymers, aerospace alloys, solid electrolytes, biomedical materials \\
\bottomrule
\end{tabular}
}
\end{table}

\section{Benchmark Algorithm Details}
\label{app:algorithm_details}

\subsection{Unified target-conditioned proposal interface}

All algorithms are evaluated through the same target-conditioned proposal interface. At round \(t\),
an algorithm observes \(D_t=\{(x_i,f(x_i))\}_{i=1}^{n_t}\), the design space \(\mathcal{X}\), and the
target specification \(y^\star\). It then induces a proposal distribution
\begin{equation}
S^{(t)}\sim \pi_{\mathcal{A}}(\cdot\mid D_t,y^\star),
\label{eq:app_proposal_policy}
\end{equation}
where \(S^{(t)}=\{x_1^{(t)},\ldots,x_K^{(t)}\}\) is the submitted candidate set. Only these submitted
candidates are evaluated by the clean oracle. This design prevents algorithms that internally generate
large candidate pools from receiving a larger oracle budget.

For methods that generate an internal pool \(C_{\mathcal{A}}=\{\tilde{x}_1,\ldots,\tilde{x}_{N}\}\),
MatFormBench ranks candidates using a target-satisfaction utility before submitting the top \(K\)
candidates. Given a surrogate prediction \(\hat{f}(x)\), a generic utility can be written as
\begin{equation}
u(x;y^\star)
=
\sum_{j=1}^{M}w_j\mathbf{1}\{c_j(\hat{f}(x),y^\star)\leq 0\}
-
\lambda\sum_{j=1}^{M}\left[c_j(\hat{f}(x),y^\star)\right]_+,
\label{eq:app_target_utility}
\end{equation}
where \(c_j\) denotes the violation of the \(j\)-th target condition and \([a]_+=\max(a,0)\).
This utility rewards target hits and penalizes constraint violations. Final benchmark scores are
always computed using the clean oracle rather than this internal utility.

\subsection{Surrogate models}

Surrogate models approximate the response map \(f:\mathcal{X}\to\mathbb{R}^3\) from observed
data. Given \(D=\{(x_i,y_i)\}_{i=1}^{n}\), a surrogate \(\hat{f}_{\theta}\) is trained by minimizing
\begin{equation}
\min_{\theta}
\sum_{i=1}^{n}\|\hat{f}_{\theta}(x_i)-y_i\|_2^2+\Omega(\theta),
\label{eq:app_surrogate_objective}
\end{equation}
where \(\Omega(\theta)\) is a model-dependent regularizer. Ridge regression uses an \(L_2\)-regularized
linear estimator and provides a stable low-variance baseline \cite{hoerl1970ridge}. Support vector
regression uses an \(\epsilon\)-insensitive loss with margin-based regularization \cite{cortes1995svm}.
Gaussian process regression places a prior over functions,
\begin{equation}
f(x)\sim \mathcal{GP}(m(x),k(x,x')),
\label{eq:app_gp_prior}
\end{equation}
and returns both a posterior mean and uncertainty estimate \cite{rasmussen2006gp}. In MatFormBench,
surrogates are used for internal ranking, Bayesian acquisition, or heuristic search, while all submitted
candidates are evaluated by the clean oracle.

\subsection{Diffusion-based generative models}

Diffusion models learn to reverse a gradual noising process. Given a clean formulation \(x_0\), the
forward process samples
\begin{equation}
q(x_t\mid x_0)
=
\mathcal{N}\left(\sqrt{\bar{\alpha}_t}x_0,(1-\bar{\alpha}_t)I\right),
\label{eq:app_diffusion_forward}
\end{equation}
where \(\bar{\alpha}_t\) is determined by the noise schedule. A denoising network
\(\epsilon_{\theta}(x_t,t,y^\star)\) is trained with
\begin{equation}
\mathcal{L}_{\mathrm{DDPM}}
=
\mathbb{E}_{x_0,t,\epsilon}
\left[
\|\epsilon-\epsilon_{\theta}(x_t,t,y^\star)\|_2^2
\right].
\label{eq:app_ddpm_loss}
\end{equation}
DDPM samples candidates by iteratively applying a stochastic reverse process \cite{ho2020denoising}.
DDIM uses a deterministic non-Markovian sampler to reduce sampling cost \cite{song2020ddim}.
EDM improves diffusion sampling through noise-level parameterization and preconditioning
\cite{karras2022edm}. Flow Matching instead learns a continuous vector field
\(v_{\theta}(x_t,t,y^\star)\) satisfying
\begin{equation}
\frac{dx_t}{dt}=v_{\theta}(x_t,t,y^\star),
\label{eq:app_flow_matching_ode}
\end{equation}
and generates candidates by integrating this learned flow \cite{lipman2022flow}.

In MatFormBench, diffusion models are used as conditional formulation generators. They approximate
\(p_{\theta}(x\mid y^\star,D_t)\), generate a large candidate pool, and rank candidates by the
target-satisfaction utility before clean-oracle validation. This mechanism is suitable for multimodal
and sparse-feasible landscapes because stochastic sampling can cover separated regions while
conditioning biases candidates toward the target.

\subsection{Variational autoencoder models}

Variational autoencoders define a latent-variable model
\begin{equation}
p_{\theta}(x,z\mid y^\star)=p(z)p_{\theta}(x\mid z,y^\star),
\label{eq:app_vae_joint}
\end{equation}
with approximate posterior \(q_{\phi}(z\mid x,y^\star)\). The standard VAE objective maximizes
the evidence lower bound
\begin{equation}
\mathcal{L}_{\mathrm{VAE}}
=
\mathbb{E}_{q_{\phi}(z\mid x)}
[\log p_{\theta}(x\mid z)]
-
D_{\mathrm{KL}}(q_{\phi}(z\mid x)\|p(z))
\label{eq:app_vae_elbo}
\end{equation}
\cite{kingma2014auto}. BetaVAE introduces a weighted KL penalty,
\begin{equation}
\mathcal{L}_{\beta\mathrm{VAE}}
=
\mathbb{E}_{q_{\phi}(z\mid x)}
[\log p_{\theta}(x\mid z)]
-
\beta D_{\mathrm{KL}}(q_{\phi}(z\mid x)\|p(z)),
\label{eq:app_betavae_objective}
\end{equation}
which encourages stronger latent regularization \cite{higgins2017betavae}. CVAE conditions both
the encoder and decoder on \(y^\star\), enabling target-aware generation \cite{sohn2015cvae}. IWAE
uses multiple importance-weighted samples and optimizes
\begin{equation}
\mathcal{L}_{\mathrm{IWAE}}
=
\mathbb{E}_{z_1,\ldots,z_K}
\left[
\log
\frac{1}{K}
\sum_{k=1}^{K}
\frac{p_{\theta}(x,z_k)}
{q_{\phi}(z_k\mid x)}
\right],
\label{eq:app_iwae_objective}
\end{equation}
which gives a tighter variational bound \cite{burda2016iwae}. VampPrior replaces the standard
Gaussian prior by a learned mixture of variational posteriors,
\begin{equation}
p(z)=\frac{1}{M}\sum_{m=1}^{M}q_{\phi}(z\mid u_m),
\label{eq:app_vampprior}
\end{equation}
where \(u_m\) are pseudo-inputs \cite{tomczak2018vamprior}.

In MatFormBench, VAE models sample latent variables and decode them into formulation candidates.
They are effective when feasible regions lie on relatively smooth latent manifolds. However, when
the feasible set becomes discontinuous or globally sparse, latent interpolation can generate invalid
or off-target candidates.

\subsection{Generative adversarial models}

Generative adversarial networks optimize a minimax game between a generator \(G_{\theta}\) and
a discriminator \(D_{\psi}\):
\begin{equation}
\min_G\max_D
\mathbb{E}_{x\sim p_{\mathrm{data}}}[\log D(x)]
+
\mathbb{E}_{z\sim p(z)}[\log(1-D(G(z)))]
\label{eq:app_gan_objective}
\end{equation}
\cite{goodfellow2014generative}. CGAN conditions both generator and discriminator on target
information, using \(G(z,y^\star)\) and \(D(x,y^\star)\) \cite{mirza2014cgan}. WGAN-GP replaces
the original adversarial objective with a Wasserstein objective and gradient penalty,
\begin{equation}
\mathcal{L}_{\mathrm{WGAN-GP}}
=
\mathbb{E}_{\tilde{x}}[D(\tilde{x})]
-
\mathbb{E}_{x}[D(x)]
+
\lambda
\mathbb{E}_{\hat{x}}
\left[
(\|\nabla_{\hat{x}}D(\hat{x})\|_2-1)^2
\right],
\label{eq:app_wgangp_objective}
\end{equation}
which improves stability \cite{gulrajani2017wgangp}. PacGAN packs multiple samples into the
discriminator to reduce mode collapse \cite{lin2018pacgan}. CTGAN adapts adversarial generation
to tabular data using conditional sampling and mode-specific normalization \cite{xu2019ctgan}.

In MatFormBench, GANs generate implicit candidate distributions and are followed by surrogate-based
candidate ranking. They can provide diverse candidates, but adversarial training can be unstable in
small-data regimes and does not directly optimize target satisfaction.

\subsection{Bayesian optimization methods}

Bayesian optimization treats inverse design as expensive black-box optimization. A probabilistic
surrogate provides posterior mean \(\mu_t(x)\) and uncertainty \(\sigma_t(x)\), and candidates are
selected by maximizing an acquisition function,
\begin{equation}
x_{t+1}=\arg\max_{x\in\mathcal{X}}a_t(x)
\label{eq:app_bo_policy}
\end{equation}
\cite{shahriari2016bayesian,frazier2018tutorial}. RGP-UCB uses an upper-confidence-bound
criterion,
\begin{equation}
a_{\mathrm{UCB}}(x)=\mu_t(x)+\kappa\sigma_t(x),
\label{eq:app_ucb_acquisition}
\end{equation}
where \(\kappa\) controls exploration \cite{srinivas2010gpucb}. TargetEGO follows efficient global
optimization and uses expected improvement,
\begin{equation}
a_{\mathrm{EI}}(x)
=
\mathbb{E}[\max(u(x)-u^+,0)],
\label{eq:app_ei_acquisition}
\end{equation}
where \(u^+\) is the best observed utility \cite{jones1998ego}. GoalPA uses a target-probability
criterion,
\begin{equation}
a_{\mathrm{goal}}(x)
=
\Pr(f(x)\in \mathcal{T}(y^\star)\mid D_t),
\label{eq:app_goal_acquisition}
\end{equation}
which directly prioritizes candidates likely to satisfy the target region.

In MatFormBench, Bayesian optimization is implemented over sampled candidate pools because the
design space is continuous and moderately high-dimensional. BO is effective when uncertainty is
well calibrated, but it can struggle when feasible target regions are rare or separated by invalid
regions.

\subsection{Surrogate-assisted heuristic search}

Surrogate-assisted heuristic search approximately solves
\begin{equation}
x^\star\approx \arg\max_{x\in\mathcal{X}}\hat{u}(x;y^\star),
\label{eq:app_surrogate_search_objective}
\end{equation}
where \(\hat{u}\) is a surrogate-estimated target utility. Genetic algorithms update a population
through selection, crossover, and mutation,
\begin{equation}
P_{t+1}=
\mathrm{Mutation}(\mathrm{Crossover}(\mathrm{Selection}(P_t))),
\label{eq:app_ga_update}
\end{equation}
and have been widely used in materials discovery \cite{holland1975ga,jennings2019genetic}. Particle
swarm optimization updates particle velocities and positions according to
\begin{equation}
\begin{aligned}
v_i^{t+1}
&=
\omega v_i^t
+
c_1r_1(p_i^t-x_i^t)
+
c_2r_2(g^t-x_i^t),\\
x_i^{t+1}
&=
x_i^t+v_i^{t+1},
\end{aligned}
\label{eq:app_pso_update}
\end{equation}
where \(p_i^t\) and \(g^t\) are personal and global best candidates \cite{kennedy1995particle}.
Simulated annealing accepts a new candidate \(x'\) with probability
\begin{equation}
\min\{1,\exp((\hat{u}(x')-\hat{u}(x))/T)\},
\label{eq:app_sa_acceptance}
\end{equation}
where \(T\) is gradually decreased \cite{kirkpatrick1983optimization}. Monte Carlo search samples
candidates from the design space and ranks them by surrogate utility; tree-search variants are
related to Monte Carlo tree search methods \cite{browne2012mcts}. Grey wolf optimization uses
leader-based population updates \cite{mirjalili2014gwo}; ant colony optimization uses pheromone-like
search preferences \cite{dorigo1996ant}; and the firefly algorithm moves candidates toward brighter
solutions, where brightness corresponds to surrogate utility \cite{yang2009firefly}.

These methods are less expressive than deep generative models, but they can be competitive under
local discontinuities because population-based search can exploit local surrogate structure and avoid
known invalid regions.

\subsection{LLM-based recommendation}

The LLM baselines are formulated as direct target-conditioned recommendation systems. Given a
prompt containing \(D_t\), design bounds, and target constraints, the model returns a structured
candidate set. Formally, the language model defines
\begin{equation}
p_{\theta}(s\mid \mathrm{prompt}(D_t,y^\star)),
\label{eq:app_llm_distribution}
\end{equation}
where the output string \(s\) must be parsed into numerical candidate vectors. A valid output must
have the correct dimensionality, contain numerical values, and satisfy the required schema.

DeepSeek-v4-flash produces valid structured candidates and is included in the reported LLM scores.
GLM-5.1 and Kimi K2.6 fail to produce valid oracle-evaluable candidates under the same protocol;
their results are therefore reported as missing. This failure mode indicates that language-model-only
recommendation does not reliably enforce numerical feasibility, dimensional consistency, or target
constraint satisfaction. Tool-augmented scientific LLM systems such as ChemCrow and Coscientist
show that LLMs can support scientific planning and recommendation \cite{bran2024chemcrow,boiko2023coscientist},
but strict continuous formulation inverse design requires numerical optimization, constrained decoding,
or tool-based validation.

\section{Implementation and Hyperparameter Settings}
\label{app:implementation}

\subsection{Common benchmark settings}

All algorithms are evaluated using the same task registry, design bounds, target constraints, and
clean validation oracle. The default initial training set contains \(n_0=30\) Latin hypercube samples.
Each non-LLM method generates an internal candidate pool and submits the top \(K=5\) candidates
for validation. LLM baselines directly submit \(K=5\) candidates. Closed-loop evaluation uses
\(R=5\) rounds, resulting in a total validation budget of \(B=25\) candidates per algorithm-task pair.

For robustness analysis, the initial training size is varied over
\begin{equation}
\mathcal{N}=\{10,15,30,50,100\}.
\label{eq:app_training_size_grid}
\end{equation}
For stability analysis, repeated runs use seeds
\begin{equation}
\mathcal{S}=\{11,22,33,44,55,66,77,88,99,111\}.
\label{eq:app_stability_seed_set}
\end{equation}
Malformed outputs, infeasible candidates, missing values, and parsing failures are uniformly treated
as failed candidates.

\subsection{Generative model settings}

\begin{table}[h]
\centering
\caption{Hyperparameters for deep generative baselines.}
\label{tab:app_generative_hyperparams}
\small
\renewcommand{\arraystretch}{1.15}
\resizebox{\linewidth}{!}{
\begin{tabular}{l l c c c c c}
\toprule
\textbf{Family} & \textbf{Variants} & \textbf{Generated pool} & \textbf{Ranked pool} & \textbf{Hidden/latent dim.} & \textbf{Epochs} & \textbf{Learning rate} \\
\midrule
Diffusion 
&
DDPM~\cite{ho2020denoising}, DDIM~\cite{song2020ddim}, EDM~\cite{karras2022edm}, Flow Matching~\cite{lipman2022flow}
& 1000 & 512 & hidden \(=96\) & 600 & \(10^{-3}\) \\
VAE 
&
BetaVAE~\cite{higgins2017betavae}, CVAE~\cite{sohn2015cvae}, IWAE~\cite{burda2016iwae}, VampPrior~\cite{tomczak2018vamprior}
& 1000 & 512 & latent \(=8\), hidden \(=64\) & 500 & \(10^{-3}\) \\
GAN 
&
CGAN~\cite{mirza2014cgan}, CTGAN~\cite{xu2019ctgan}, PacGAN~\cite{lin2018pacgan}, WGAN-GP~\cite{gulrajani2017wgangp}
& 1000 & 512 & latent \(=16\), hidden \(=64\) & 400 & \(10^{-3}\) \\
\bottomrule
\end{tabular}
}
\end{table}

All deep generative models use mini-batch training with batch size \(32\). Diffusion models use
\(50\) diffusion steps by default. DDIM uses \(20\) sampling steps, while EDM and Flow Matching
use \(32\) sampling steps. Generated candidates are ranked by the internal target utility before the
top \(K=5\) candidates are submitted to the clean oracle.

\subsection{Surrogate-assisted search settings}

\begin{table}[h]
\centering
\caption{Hyperparameters for surrogate-assisted heuristic optimizers.}
\label{tab:app_search_hyperparams}
\small
\renewcommand{\arraystretch}{1.15}
\begin{tabular}{l l}
\toprule
\textbf{Optimizer} & \textbf{Configuration} \\
\midrule
Genetic algorithm~\cite{holland1975ga,jennings2019genetic} & population-based selection, crossover, and mutation \\
Particle swarm optimization~\cite{kennedy1995particle} & 100 particles, 200 iterations \\
Monte Carlo search~\cite{browne2012mcts} & 5000 sampled candidates \\
Simulated annealing~\cite{kirkpatrick1983optimization} & 1000 iterations \\
Grey wolf optimization~\cite{mirjalili2014gwo} & 200 iterations \\
Ant colony optimization~\cite{dorigo1996ant} & 80 ants, 200 iterations, evaporation rate 0.2 \\
Firefly algorithm~\cite{yang2009firefly} & 200 iterations, \(\alpha=0.2\) \\
\bottomrule
\end{tabular}
\end{table}

\begin{table}[h]
\centering
\caption{Surrogate model configurations.}
\label{tab:app_surrogate_hyperparams}
\small
\renewcommand{\arraystretch}{1.15}
\begin{tabular}{l l}
\toprule
\textbf{Surrogate} & \textbf{Configuration} \\
\midrule
Ridge regression~\cite{hoerl1970ridge} & default ridge model \\
Gaussian process regression~\cite{rasmussen2006gp} & normalized targets, trend-Matern kernel, one optimizer restart \\
Support vector regression~\cite{cortes1995svm} & linear kernel, \(C=10.0\), \(\epsilon=0.01\), \(\gamma=\mathrm{scale}\) \\
\bottomrule
\end{tabular}
\end{table}

\subsection{LLM settings}

\begin{table}[h]
\centering
\caption{LLM recommendation settings. GLM-5.1 and Kimi K2.6 did not produce valid oracle-evaluable outputs and are excluded from aggregate LLM scores.}
\label{tab:app_llm_settings}
\small
\renewcommand{\arraystretch}{1.15}
\begin{tabular}{l l c c}
\toprule
\textbf{Model} & \textbf{Execution mode} & \textbf{Valid output} & \textbf{Included in score} \\
\midrule
DeepSeek-v4-flash~\cite{deepseek2026v4} & API-based direct recommendation & Yes & Yes \\
GLM-5.1~\cite{glm2026technical} & API-based direct recommendation & No & No \\
Kimi K2.6~\cite{kimi2026technical} & API-based direct recommendation & No & No \\
\bottomrule
\end{tabular}
\end{table}

\section{Computational Resources}
\label{app:compute}

All non-API experiments were executed on Google Cloud Platform. The main compute worker used
one NVIDIA A100 GPU with \(40\) GB memory. The CPU configuration was Intel Cascade Lake with
\(12\) vCPUs and \(85\) GB system memory. Deep generative models were executed on the GPU when
CUDA was available, while surrogate-assisted search and Bayesian optimization primarily used CPU
resources. The LLM baselines were evaluated through external API calls using API keys.

\begin{table}[h]
\centering
\caption{Compute environment used for the benchmark experiments.}
\label{tab:app_compute}
\small
\renewcommand{\arraystretch}{1.15}
\begin{tabular}{l l}
\toprule
\textbf{Resource} & \textbf{Configuration} \\
\midrule
Cloud provider & Google Cloud Platform \\
GPU & 1 \(\times\) NVIDIA A100, 40 GB \\
CPU & Intel Cascade Lake \\
vCPUs & 12 \\
System memory & 85 GB \\
LLM execution & External API calls using API keys \\
Main software stack & Python, NumPy, pandas, scikit-learn, PyTorch \\
Validation oracle & Local clean oracle evaluation \\
\bottomrule
\end{tabular}
\end{table}

Because LLM baselines rely on external APIs, their runtime may depend on provider-side latency,
rate limits, and model availability. However, all returned candidates are evaluated locally under the
same clean-oracle protocol as the non-LLM methods.

\section{Supplementary Experimental Results}
\label{app:supp_results}

\subsection{Family-level MatFormScore with variability}

Table~\ref{tab:app_family_level_score} reports family-level MatFormScore across the five difficulty
levels. Values are reported as mean \(\pm\) standard deviation over the corresponding algorithm-task
instances.

\begin{table}[h]
\centering
\caption{Family-level MatFormScore across difficulty levels. Values are mean \(\pm\) standard deviation.}
\label{tab:app_family_level_score}
\scriptsize
\renewcommand{\arraystretch}{1.15}
\resizebox{\linewidth}{!}{
\begin{tabular}{l c c c c c}
\toprule
\textbf{Family} & \textbf{L1} & \textbf{L2} & \textbf{L3} & \textbf{L4} & \textbf{L5} \\
\midrule
Diffusion & \(54.8\pm22.9\) & \(68.3\pm19.0\) & \(55.4\pm21.4\) & \(72.5\pm11.2\) & \(61.8\pm12.3\) \\
VAE & \(59.1\pm21.8\) & \(58.3\pm19.5\) & \(39.7\pm14.8\) & \(60.6\pm17.7\) & \(55.4\pm17.6\) \\
GAN & \(54.2\pm20.6\) & \(61.3\pm20.0\) & \(48.5\pm22.9\) & \(66.7\pm11.3\) & \(48.2\pm20.1\) \\
LLM & \(41.7\pm12.9\) & \(31.7\pm4.1\) & \(43.2\pm14.9\) & \(41.3\pm11.6\) & \(42.3\pm12.8\) \\
Search & \(37.0\pm20.3\) & \(41.4\pm19.8\) & \(41.2\pm18.4\) & \(46.1\pm16.4\) & \(25.8\pm19.0\) \\
Bayesian Optimization & \(33.3\pm22.9\) & \(40.1\pm18.1\) & \(38.4\pm18.0\) & \(33.8\pm13.5\) & \(22.5\pm13.6\) \\
\bottomrule
\end{tabular}
}
\end{table}

\begin{figure}[h]
\centering
\includegraphics[width=0.82\linewidth]{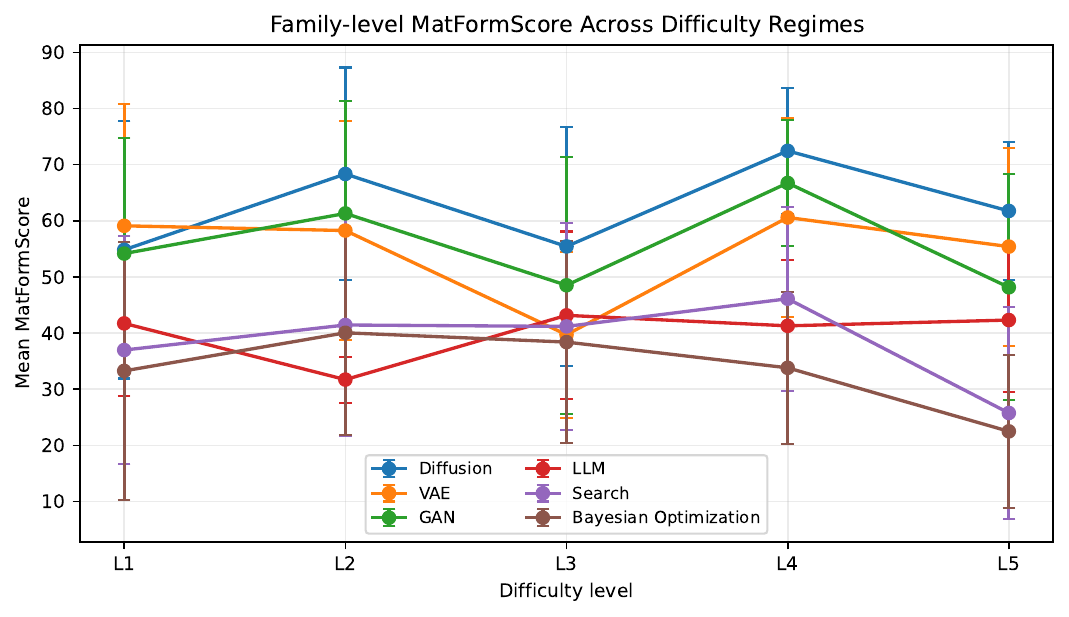}
\caption{Family-level MatFormScore across difficulty levels with error bars. Points denote mean MatFormScore and error bars denote one standard deviation over the corresponding algorithm-task instances.}
\label{fig:app_family_errorbar}
\end{figure}

\subsection{Full algorithm ranking}

Table~\ref{tab:app_full_ranking} reports the full algorithm-level ranking over all \(30\) benchmark
datasets. GLM-5.1 and Kimi K2.6 are shown as missing because they did not produce valid
oracle-evaluable candidates under the benchmark protocol.

\begin{longtable}{c l l c}
\caption{Full algorithm ranking by mean MatFormScore. Values are mean \(\pm\) standard deviation over benchmark datasets.}
\label{tab:app_full_ranking}\\
\toprule
\textbf{Rank} & \textbf{Algorithm} & \textbf{Family} & \textbf{MatFormScore} \\
\midrule
\endfirsthead
\toprule
\textbf{Rank} & \textbf{Algorithm} & \textbf{Family} & \textbf{MatFormScore} \\
\midrule
\endhead
1 & DDPM~\cite{ho2020denoising} & Diffusion & \(65.78\pm21.17\) \\
2 & DDIM~\cite{song2020ddim} & Diffusion & \(65.54\pm19.77\) \\
3 & CVAE~\cite{sohn2015cvae} & VAE & \(63.29\pm17.61\) \\
4 & IWAE~\cite{burda2016iwae} & VAE & \(61.96\pm17.91\) \\
5 & WGAN-GP~\cite{gulrajani2017wgangp} & GAN & \(61.16\pm20.26\) \\
6 & EDM~\cite{karras2022edm} & Diffusion & \(59.49\pm16.21\) \\
7 & Flow Matching~\cite{lipman2022flow} & Diffusion & \(59.43\pm18.66\) \\
8 & CTGAN~\cite{xu2019ctgan} & GAN & \(58.27\pm18.21\) \\
9 & GA-SVR~\cite{holland1975ga,cortes1995svm} & Search & \(56.89\pm26.85\) \\
10 & GA-Ridge~\cite{holland1975ga,hoerl1970ridge} & Search & \(55.71\pm26.29\) \\
11 & CGAN~\cite{mirza2014cgan} & GAN & \(53.26\pm23.22\) \\
12 & GA-GPR~\cite{holland1975ga,rasmussen2006gp} & Search & \(50.72\pm20.89\) \\
13 & PacGAN~\cite{lin2018pacgan} & GAN & \(50.44\pm19.14\) \\
14 & VampPrior~\cite{tomczak2018vamprior} & VAE & \(49.52\pm17.27\) \\
15 & Wolf-SVR~\cite{mirjalili2014gwo,cortes1995svm} & Search & \(43.81\pm20.57\) \\
16 & BetaVAE~\cite{higgins2017betavae} & VAE & \(43.69\pm19.61\) \\
17 & Wolf-Ridge~\cite{mirjalili2014gwo,hoerl1970ridge} & Search & \(40.94\pm19.04\) \\
18 & GoalPA~\cite{shahriari2016bayesian} & Bayesian Optimization & \(40.06\pm20.09\) \\
19 & DeepSeek-v4-flash~\cite{deepseek2026v4} & LLM & \(40.04\pm12.60\) \\
20 & MC-Ridge~\cite{browne2012mcts,hoerl1970ridge} & Search & \(37.93\pm18.08\) \\
21 & Firefly-Ridge~\cite{yang2009firefly,hoerl1970ridge} & Search & \(37.90\pm16.79\) \\
22 & Firefly-SVR~\cite{yang2009firefly,cortes1995svm} & Search & \(37.68\pm17.45\) \\
23 & MC-SVR~\cite{browne2012mcts,cortes1995svm} & Search & \(37.45\pm18.30\) \\
24 & PSO-Ridge~\cite{kennedy1995particle,hoerl1970ridge} & Search & \(36.88\pm19.30\) \\
25 & SA-SVR~\cite{kirkpatrick1983optimization,cortes1995svm} & Search & \(36.26\pm17.99\) \\
26 & Firefly-GPR~\cite{yang2009firefly,rasmussen2006gp} & Search & \(35.64\pm13.35\) \\
27 & PSO-SVR~\cite{kennedy1995particle,cortes1995svm} & Search & \(34.76\pm18.36\) \\
28 & MC-GPR~\cite{browne2012mcts,rasmussen2006gp} & Search & \(34.64\pm16.26\) \\
29 & PSO-GPR~\cite{kennedy1995particle,rasmussen2006gp} & Search & \(34.58\pm15.93\) \\
30 & RGP-UCB~\cite{srinivas2010gpucb} & Bayesian Optimization & \(34.28\pm17.18\) \\
31 & SA-Ridge~\cite{kirkpatrick1983optimization,hoerl1970ridge} & Search & \(33.97\pm19.06\) \\
32 & Wolf-GPR~\cite{mirjalili2014gwo,rasmussen2006gp} & Search & \(33.26\pm19.26\) \\
33 & ACO-Ridge~\cite{dorigo1996ant,hoerl1970ridge} & Search & \(32.63\pm15.81\) \\
34 & SA-GPR~\cite{kirkpatrick1983optimization,rasmussen2006gp} & Search & \(31.76\pm12.45\) \\
35 & ACO-GPR~\cite{dorigo1996ant,rasmussen2006gp} & Search & \(31.61\pm17.41\) \\
36 & ACO-SVR~\cite{dorigo1996ant,cortes1995svm} & Search & \(29.26\pm16.09\) \\
37 & TargetEGO~\cite{jones1998ego} & Bayesian Optimization & \(26.48\pm15.79\) \\
38 & GLM-5.1~\cite{glm2026technical} & LLM & -- \\
39 & Kimi K2.6~\cite{kimi2026technical} & LLM & -- \\
\bottomrule
\end{longtable}

\begin{figure}[h]
\centering
\includegraphics[width=0.9\linewidth]{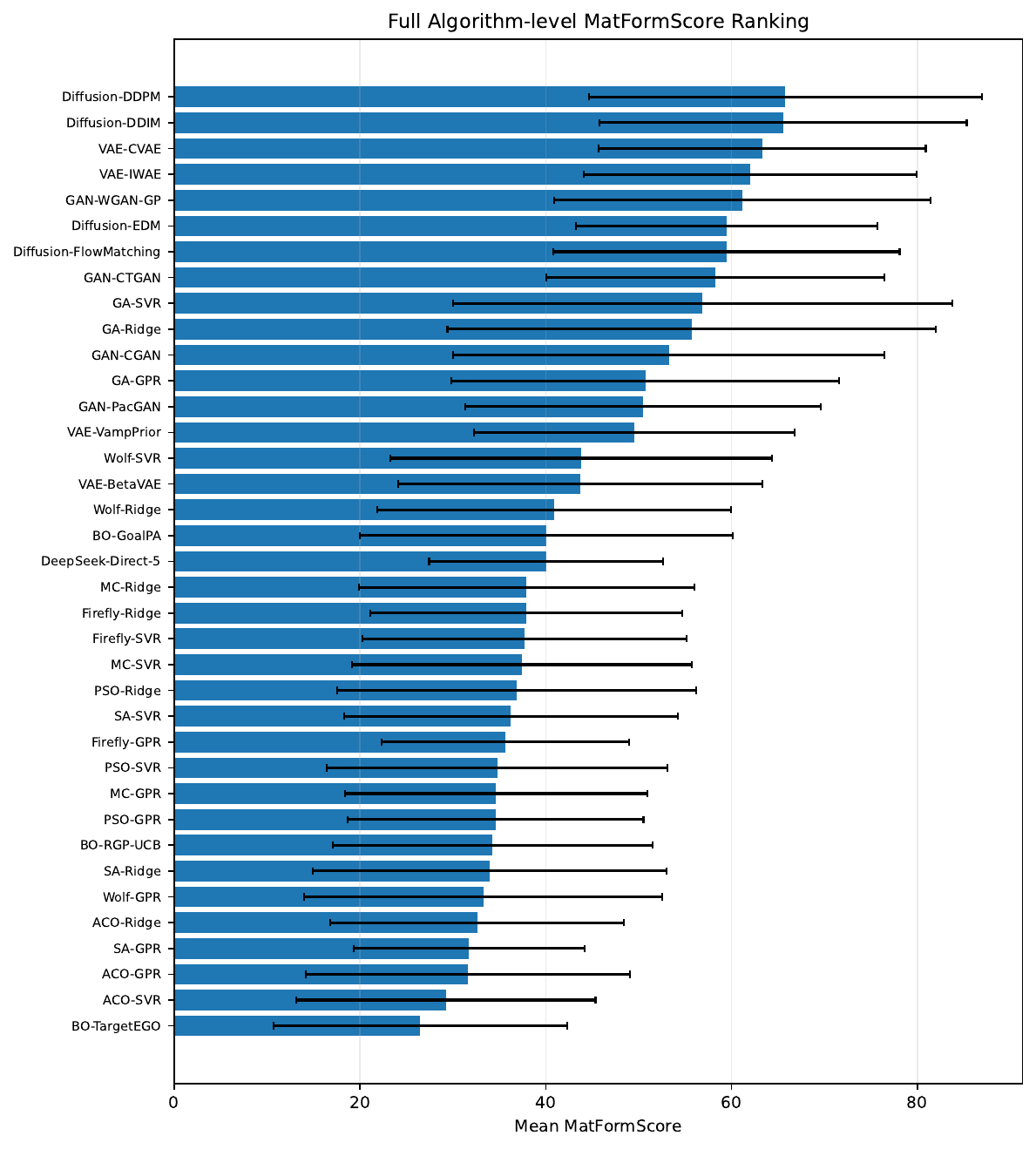}
\caption{Full algorithm-level MatFormScore ranking with error bars. Bars indicate mean MatFormScore and error bars indicate one standard deviation over benchmark datasets.}
\label{fig:app_algorithm_ranking}
\end{figure}

\subsection{Robustness across training-set sizes}

Table~\ref{tab:app_robustness} reports all-target hit rate under different initial training sizes.
Values are mean \(\pm\) standard deviation over the corresponding runs.

\begin{table}[h]
\centering
\caption{Robustness across initial training-set sizes. Values are all-target hit rate, reported as mean \(\pm\) standard deviation.}
\label{tab:app_robustness}
\scriptsize
\renewcommand{\arraystretch}{1.15}
\resizebox{\linewidth}{!}{
\begin{tabular}{l c c c c c}
\toprule
\textbf{Family} & \(n=10\) & \(n=15\) & \(n=30\) & \(n=50\) & \(n=100\) \\
\midrule
Diffusion & \(0.352\pm0.342\) & \(0.473\pm0.331\) & \(0.398\pm0.348\) & \(0.508\pm0.301\) & \(0.583\pm0.335\) \\
VAE & \(0.270\pm0.373\) & \(0.277\pm0.378\) & \(0.255\pm0.368\) & \(0.350\pm0.377\) & \(0.340\pm0.372\) \\
GAN & \(0.297\pm0.342\) & \(0.335\pm0.345\) & \(0.330\pm0.367\) & \(0.388\pm0.335\) & \(0.472\pm0.324\) \\
LLM & \(0.133\pm0.166\) & \(0.073\pm0.150\) & \(0.180\pm0.202\) & \(0.153\pm0.211\) & \(0.100\pm0.169\) \\
Search & \(0.203\pm0.285\) & \(0.201\pm0.305\) & \(0.123\pm0.264\) & \(0.074\pm0.197\) & \(0.078\pm0.200\) \\
Bayesian Optimization & \(0.156\pm0.219\) & \(0.156\pm0.256\) & \(0.116\pm0.221\) & \(0.071\pm0.147\) & \(0.056\pm0.119\) \\
\bottomrule
\end{tabular}
}
\end{table}

\begin{figure}[h]
\centering
\includegraphics[width=0.82\linewidth]{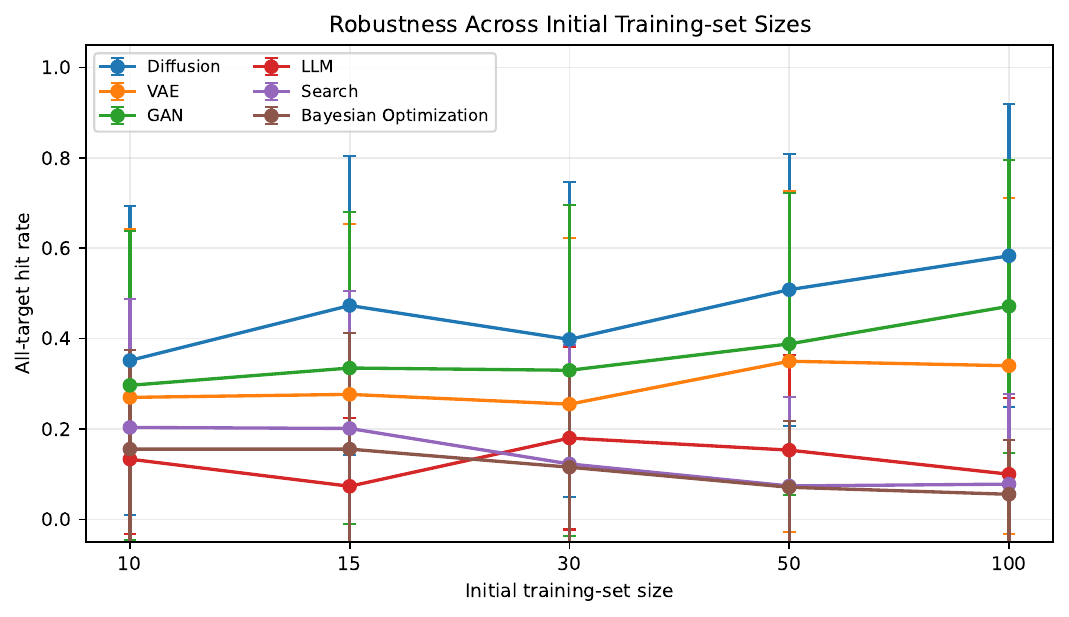}
\caption{Robustness under varying initial training-set sizes. Points denote mean all-target hit rate and error bars denote one standard deviation.}
\label{fig:app_robustness}
\end{figure}

\subsection{Seed-level stability}

Table~\ref{tab:app_stability} reports seed-level all-target hit-rate variability grouped by algorithm
family.

\begin{table}[h]
\centering
\caption{Seed-level stability by algorithm family. Values are mean \(\pm\) standard deviation of all-target hit rate.}
\label{tab:app_stability}
\small
\renewcommand{\arraystretch}{1.15}
\begin{tabular}{l c}
\toprule
\textbf{Family} & \textbf{Seed-level all-target hit rate} \\
\midrule
Diffusion & \(0.453\pm0.211\) \\
GAN & \(0.352\pm0.222\) \\
VAE & \(0.306\pm0.239\) \\
LLM & \(0.151\pm0.075\) \\
Bayesian Optimization & \(0.126\pm0.139\) \\
Search & \(0.122\pm0.181\) \\
\bottomrule
\end{tabular}
\end{table}

\begin{figure}[h]
\centering
\includegraphics[width=0.75\linewidth]{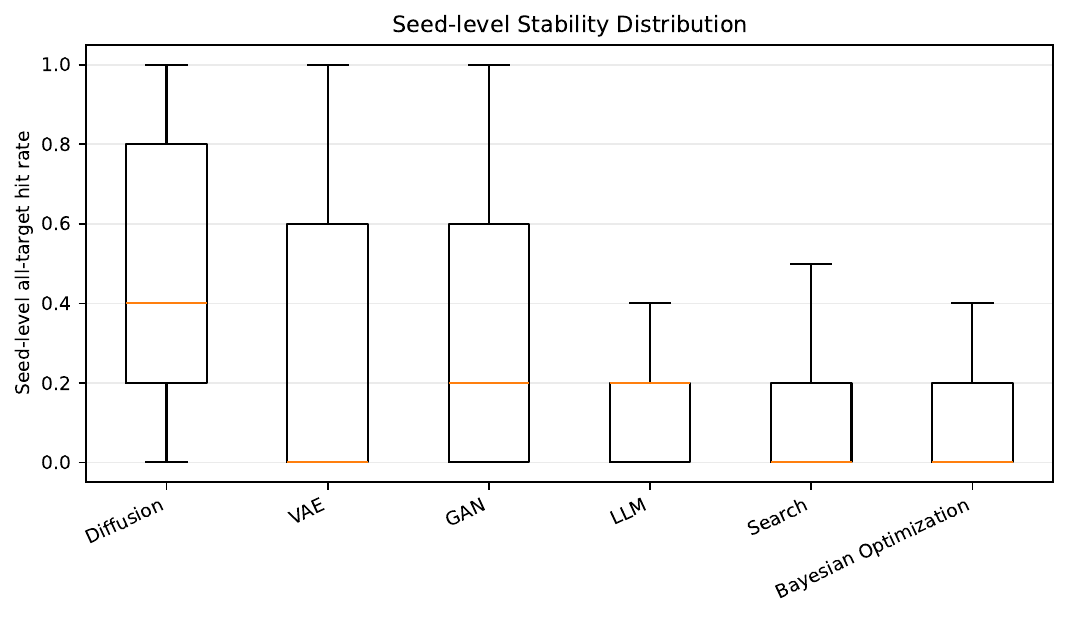}
\caption{Seed-level stability distribution across algorithm families. The distribution summarizes all-target hit rates across repeated random seeds.}
\label{fig:app_seed_stability}
\end{figure}

\subsection{LLM failure cases}

\begin{table}[h]
\centering
\caption{LLM baseline status. GLM-5.1 and Kimi K2.6 are reported as missing because they failed to generate valid oracle-evaluable structured candidates.}
\label{tab:app_llm_failure}
\small
\renewcommand{\arraystretch}{1.15}
\resizebox{\linewidth}{!}{
\begin{tabular}{l l c c c}
\toprule
\textbf{Model} & \textbf{Execution mode} & \textbf{Valid structured output} & \textbf{Included in MatFormScore} & \textbf{Result} \\
\midrule
DeepSeek-v4-flash~\cite{deepseek2026v4} & API direct recommendation & Yes & Yes & \(40.04\pm12.60\) \\
GLM-5.1~\cite{glm2026technical} & API direct recommendation & No & No & -- \\
Kimi K2.6~\cite{kimi2026technical} & API direct recommendation & No & No & -- \\
\bottomrule
\end{tabular}
}
\end{table}

These failures indicate that language-model-only recommendation is not a reliable standalone solver
for continuous formulation inverse design. Without constrained numerical decoding or tool-based
validation, LLM outputs may violate schema requirements, dimensional consistency, design bounds,
or target constraints.

\section{Reproducibility and Released Assets}
\label{app:reproducibility}

The released MatFormBench package contains the synthetic oracle, task registry, algorithm cards,
benchmark runners, metric computation scripts, and JSON output records. Each output record stores
the algorithm name, task level, dataset identifier, submitted candidates, clean oracle values, hit-rate
statistics, diversity, hypervolume, robustness statistics, stability statistics, and final MatFormScore.

Code and project page are available at
\url{https://github.com/DeepVerse/MatFormBench}
and
\url{https://matformbench.deepverse.tech}.

The benchmark fixes the task registry, design bounds, target constraints, Latin hypercube sampling
rule, validation oracle, search budget, number of submitted candidates, random seeds, and failure
handling rules. Therefore, each reported result can be traced back to an algorithm-task evaluation
record and recomputed from the released outputs.

For non-LLM algorithms, the released code and output files are sufficient to reproduce the benchmark
tables and figures. For LLM baselines, exact reruns may depend on provider-side model availability,
API versioning, and stochastic decoding behavior. Accordingly, the released JSON records should be
treated as the reference outputs for the reported LLM experiments. DeepSeek-v4-flash produced valid
structured candidates, while GLM-5.1 and Kimi K2.6 did not produce valid oracle-evaluable outputs
under the required protocol.

\FloatBarrier



\end{document}